\documentclass[floatfix,aps,twocolumn,preprintnumbers,amsmath,amssymb,nofootinbib,superscriptaddress]{revtex4}

\usepackage{graphicx,color}
\usepackage{amsmath,amssymb,bm}
\usepackage{dcolumn}
\usepackage{comment}
\usepackage{slashed}

\newcommand{\bea}{\begin{eqnarray}}
\newcommand{\eea}{\end{eqnarray}}
\newcommand{\be}{\begin{equation}}
\newcommand{\ee}{\end{equation}}
\newcommand{\np}{{\bf p}}

\newcommand{\nh}{{\bf h}}

\newcommand{\nk}{{\bf k}}

\newcommand{\nq}{{\bf q}}

\newcommand{\nj}{{\bf j}}

\newcommand{\nK}{{\bf K}}
\newcommand{\nA}{{\bf A}}
\newcommand{\nB}{{\bf B}}

\newcommand{\kbar}{\not{\!k}}
\newcommand{\Pbar}{\not{\!P}}
\newcommand{\pbar}{\not{\!p}}

\newcommand{\nsigma}{\mbox{\boldmath $\sigma$}}
\newcommand{\ntau}{\mbox{\boldmath $\tau$}}

\def\XXint#1#2#3{{\setbox0=\hbox{$#1{#2#3}{\int}$}
     \vcenter{\hbox{$#2#3$}}\kern-.5\wd0}}

\def\1{\'{\i}}

\begin{document}

\title{ Meson-exchange currents in quasielastic charged-current
  neutrino reactions with single-nucleon nnockout}

\author{P.R. Casale} \email{palomacasale@ugr.es}
\affiliation{Departamento de
  F\'{\i}sica At\'omica, Molecular y Nuclear}
\affiliation{Instituto Carlos I
  de F{\'\i}sica Te\'orica y Computacional Universidad de Granada,
  E-18071 Granada, Spain.}

\author{J.E. Amaro}\email{amaro@ugr.es} 
\affiliation{Departamento de
  F\'{\i}sica At\'omica, Molecular y Nuclear}
\affiliation{Instituto Carlos I
  de F{\'\i}sica Te\'orica y Computacional Universidad de Granada,
  E-18071 Granada, Spain.}

\author{V. Belocchi}%\email{belocchi@to.infn.it}
\affiliation{Dipartimento di Fisica Universit\`a di Torino, P. Giuria
  1, 10125 Torino, Italy}
\affiliation{INFN Sezione di Torino, 10125
  Torino, Italy}
\affiliation{Instituto de F\'isica Corpuscular
  (IFIC), Consejo Superior de Investigaciones Cient\'ificas (CSIC) and
  Universidad de Valencia, E-46980 Paterna, Valencia, Spain}

\author{M.B. Barbaro}%\email{maria.barbaro@unito.it} 
\affiliation{Dipartimento di Fisica Universit\`a di Torino, P. Giuria 1, 
10125 Torino, Italy}
\affiliation{INFN Sezione di Torino, 10125 Torino, Italy}

\author{M. Martini}%\email{marco.martini@ipsa.fr} 
  \affiliation{
IPSA-DRII, 63 boulevard de Brandebourg, 94200 Ivry-sur-Seine, France}
\affiliation{
Sorbonne Universit\'e, CNRS/IN2P3,\\
Laboratoire de Physique Nucl\'eaire et de Hautes Energies (LPNHE), 75005 Paris, France
}

\date{\today}

\begin{abstract}

The effect of meson-exchange currents on charged-current quasielastic
neutrino scattering with single-nucleon emission is computed and
analyzed within the relativistic Fermi gas model. This contribution
arises primarily from the interference between one-body and two-body
currents, where the two-body operator excites a 1p1h state in the
presence of a second, spectator nucleon. The results obtained show a
reduction of the vector, axial and vector-axial transverse response
functions and, consequently, a decrease in the total neutrino cross
section. In addition to a comparison with the non-relativistic limit,
other models are also explored, such as the relativistic mean field
model for nuclear matter and the superscaling analysis with
relativistic effective mass, both of which yield
qualitatively similar results.
\end{abstract}

%\pacs{24.10.Jv, 25.30.-c, 21.30.Fe, 25.30.Fj} 
%\pacs{03.65.Nk,11.10.Gh,13.75.Cs,21.30.Fe,21.45.-v}

\keywords{Quasielastic neutrino scattering, Meson-exchange currents, 
Relativistic Fermi gas, Relativistic Mean Field, Superscaling}

\maketitle

\section{Introduction}

The increasing precision of current and future neutrino experiments
demands accurate modeling of neutrino-nucleus interactions to extract
reliable information about fundamental neutrino properties, such as
masses and CP violation \cite{Abe2018,T2K2019,Dune2020}. Providing
robust theoretical predictions for cross sections across a wide
energy range and for realistic nuclear targets is a major challenge
\cite{Ank22,Sob25,Mos16,Kat16,NuSTEC17,Martini2010}. This requires combining
nuclear physics tools—effective theories, many-body methods, and
phenomenological models—to capture the complex nuclear dynamics
\cite{Pan23,Ath22,Mor12,Alv25,Ank17, Ben17,Ama20,Alv25b}.
These models are implemented in
neutrino event generators, which are essential for simulating and
analyzing experimental data \cite{Andreopoulos2010,Gallagher2011}.

In particular, a precise understanding of quasielastic (QE)
neutrino-nucleus scattering is crucial for interpreting data from
accelerator-based experiments such as MiniBooNE, T2K, MicroBooNE,
MINERvA, and the forthcoming Hyper-Kamiokande and DUNE projects \cite{Agu10,Abe2018, Aliaga2016, Abi2020}.
 The charged-current
quasielastic (CCQE) cross section is largely dominated by the
one-particle one-hole (1p1h) channel, where a single nucleon is
emitted and the residual nucleus remains in a bound state.
 However, it is now well established
that additional reaction mechanisms, including multinucleon emission
and meson-exchange currents (MEC), contribute significantly to the
total cross section and must be properly accounted for in theoretical
models to ensure a comprehensive and accurate description
\cite{Martini2009, Nie11, Nie12, Megias2016}.

In fact, an important outcome of modern theoretical developments is
the identification of an enhancement in the quasielastic-like cross
section—defined here as excluding pion production and inelastic
channels—arising from processes involving the emission of two
nucleons. These so-called two-particle–two-hole (2p2h) contributions
are associated with nuclear correlations and meson-exchange
currents. In models based on a sum over exclusive final states, the
2p2h channel corresponds to a distinct class of configurations that do
not interfere with the standard one-particle–one-hole (1p1h)
contribution, as they lead to orthogonal final states. This mechanism
has been studied using a variety of nuclear approaches, including
relativistic and non-relativistic frameworks \cite{Martini2009, Nie11,
  Megias2016, Amaro2011}, each relying on different treatments of the
nuclear dynamics and current operators. Despite methodological
differences, the predicted enhancement is relatively consistent,
typically accounting for about 20\% of the total QE cross
section. While the precise impact of this contribution on experimental
observables is still under investigation, its inclusion is necessary
for a more accurate and complete theoretical description.

The interference between MEC and the one-body current
has received limited attention in the context of neutrino
QE scattering, and remains an open issue. While the MEC
contribution has been extensively studied in the 2p2h sector, to our
knowledge, the only prior work addressing MEC in quasielastic CC
neutrino–nucleus scattering in the 1p1h channel is Ref. \cite{Umi95},
which employed a soft‑pion MEC dominance model without including the
$\Delta$ current. Apart from that, the present study represents one of
the first fully dedicated analyses of CCQE scattering in the 1p1h
channel including MEC interference effects.  In the case of inclusive
electron scattering, many investigations have highlighted the role of
the interference term, yet recent studies have raised confusion
regarding its sign \cite{Fra23,Fra25,Lov23,Cas25}.

Calculations based on independent-particle models—such as the
relativistic Fermi gas, mean-field approaches, or even spectral
function models—typically yield a negative interference in the
transverse response
\cite{Ama98,Ama02,Ama02b,Koh81,Alb90,Ama94a,Ama94b,Ama03,Cas23}.  This
behavior results from a partial cancellation between the seagull
current (which interferes positively with the one-body current) and
the $\Delta$-current and pion-in-flight diagrams (which contribute
negatively). Models such as the Valencia approach do not include the
1b–2b interference explicitly, likely under the assumption that its
net effect is small compared to other mechanisms \cite{Nieves}, such
as final-state interactions (FSI) or long-range correlations treated
within the random phase approximation (RPA). However, this assumption
remains untested.

Interestingly, a variational calculation
\cite{Fab97} showed that the sign of the interference can be
reversed depending on the type of correlations included in the nuclear
wave function: it becomes positive when tensor correlations are
incorporated via a correlated basis function approach, while it
remains negative with purely Jastrow-type correlations. This contrasts
with Green’s function Monte Carlo (GFMC) results for $^4$He
\cite{Lovato2016}, where the net effect of MEC appears to be positive
even when using simplified wave functions without tensor
correlations. Nevertheless, such comparisons must be made with
caution, as GFMC calculations cannot isolate the 1p1h channel; the
total MEC contribution includes both interference and 2p2h mechanisms,
which may partially cancel each other.  Moreover, an additional layer
of complexity arises from the interference between one-body and
two-body operators within the 2p2h channel itself \cite{Ben15}, further
complicating the disentanglement of different reaction mechanisms and
their impact on the total cross section.

Given the complexity of the problem, in this work we report the
results of a calculation performed within the simplest possible
framework, with the aim of gaining insight into the fundamental
features of the interference between one-body and two-body currents in
the 1p1h channel for CCQE neutrino scattering. This serves as a first
step towards more elaborate approaches. Specifically, we consider the
relativistic Fermi gas (RFG) model without nucleon-nucleon
correlations, and evaluate the contribution of MEC to the QE
response arising from their interference with the one-body current,
restricted to genuine 1p1h excitations, meaning final states involving only one emitted nucleon and one hole in the residual system.
The two-body current operators
are taken from a relativistic model that has been widely used in
previous studies~\cite{DePace2003, Amaro2010, Megias2016}. For
comparison, we also analyze the non-relativistic limit, in which the
vector and axial MEC reduce to the standard expressions found, for
instance, in Refs \cite{Ris89,Ericson1988}.
These currents have been extensively applied
in the context of inclusive $(e,e')$ and exclusive $(e,e'p)$
scattering.

We compute the interference responses in both the relativistic and
non-relativistic schemes, and verify that the former consistently
reduce to the latter through two independent calculations, providing a
valuable internal check. Furthermore, we explore the impact of nuclear
dynamics beyond the RFG by implementing the relativistic mean field
(RMF) model in nuclear matter \cite{Kim94,Kim96} and the super scaling
analysis with relativistic effective mass (SuSAM*) model~\cite{Cas23},
which incorporates effects associated with nuclear mean fields. We
present results for neutrino-induced interference responses and for
the flux-averaged differential cross sections relevant to
accelerator-based experiments, allowing for a preliminary comparison
with available measurements.

This paper is a natural extension to CC neutrino scattering
of Ref.~\cite{Cas25}, where the
transverse interference response was studied in electron
scattering; while avoiding unnecessary overlap by referring to that
work for the vector current formalism, we focus here on the new
ingredients, namely the axial MEC and the weak response functions,
with special attention to the axial $\Delta$ current.

In Section II we introduce
the formalism within the RFG model, along with
the relativistic MEC operators employed in
this work. In Section III, we examine the non-relativistic limit 
 and compare the results with the relativistic calculation in
for low momentum and energy transfer, where both approaches
are expected to converge. Section IV presents numerical results for the
neutrino response functions and the flux-averaged cross section,
including the interference between one-body and two-body
currents. Finally, in Section V we summarize our conclusions.

\section{Formalism}

In this section we outline the general formalism for charged-current
quasielastic (CCQE) neutrino scattering off nuclei. The framework is
based on the relativistic Fermi gas model, which allows a fully
covariant treatment of the kinematics and current operators.

\subsection{Cross section}

We follow the formalism of Ref. \cite{Ama20}. 
We consider the process \( \nu_\mu + A \rightarrow \mu^- + X
\), where a muon neutrino of four-momentum \( k^\mu = (\epsilon,
\mathbf{k}) \) scatters off a nucleus at rest, transferring
four-momentum \( Q^\mu = k^\mu - k'^\mu = (\omega, \mathbf{q}) \) to
the nuclear system. Here, \( k'^\mu = (\epsilon', \mathbf{k}') \) is
the four-momentum of the outgoing muon, and \( Q^2 = \omega^2 - q^2 <
0 \) is the squared four-momentum transfer.

The energy transfer is denoted by \( \omega = \epsilon - \epsilon' \),
and the three-momentum transfer \( \mathbf{q} \) is taken along the \(
z \)-axis. The scattering angle between the incoming neutrino and
outgoing muon is \( \theta_\mu \), so that the muon energy is \( \epsilon'
= \sqrt{|\mathbf{k}'|^2 + m_\mu^2} \), and its kinetic energy is \(
T_\mu = \epsilon' - m_\mu \).

In the laboratory frame, assuming the target nucleus is initially at
rest, the inclusive cross section reads:
\begin{equation}
\frac{d^2\sigma}{d\epsilon'd\Omega_\mu} 
= \frac{G_F^2 \cos^2\theta_C}{4\pi^2} \frac{|\mathbf{k}'|}
{\epsilon} L_{\mu\nu} W^{\mu\nu} ,
\end{equation}
where $G_F=1.666 \times 10^{-11} \;\rm MeV^{-2}$ is the Fermi constant,
\( \theta_C \) is the Cabibbo angle, \( L_{\mu\nu} \) is the leptonic
tensor, and \( W^{\mu\nu} \) is the hadronic tensor that encodes the
five nuclear response functions.  Performing the tensor contraction,
and integrating over the muon azimuthal angle, the differential
cross section with respect to $T_\mu,\cos\theta_\mu$, can be written as
\begin{eqnarray}
  \frac{d^2\sigma}{dT_\mu d\cos\theta_\mu}
&=& 
\frac{G_F^2\cos^2\theta_c}{2\pi}\frac{k'}{\epsilon}
  \frac{v_0}{2} \left( V_{CC}R_{CC}+2V_{CL}R_{CL}  \right.
\nonumber\\
&&
\left. +V_{LL}R_{LL}+V_{T} R_{T}\pm 2V_{T'} R_{T'}\right ),
\label{cross}
\end{eqnarray}
where the minus sign is for antineutrino scattering, \( \bar{\nu}_\mu + A
\rightarrow \mu^+ + X \).
In this equation we have defined the factor
$v_0 = (\epsilon+\epsilon')^2-q^2$, and
the coefficients \(V_K\) are obtained from the components of the
leptonic tensor
\begin{eqnarray}
   V_{CC}&=&1+\delta^2\frac{Q^2}{v_0},\nonumber \\  
   V_{CL}&=&\frac{\omega}{q}-\frac{\delta^2}{\rho'}\frac{Q^2}{v_0}, \nonumber \\
   V_{LL}&=&\frac{\omega^2}{q^2}+\left (1+\frac{2\omega}{q \rho'}+\rho \delta^2 \right )
   \delta^2\frac{Q^2}{v_0}, \nonumber \\
   V_{T}&=&\frac{Q^2}{v_0}+\frac{\rho}{2}-\frac{\delta^2}{\rho'}
   \left (\frac{\omega}{q}+\frac{1}{2}\rho \rho' \delta^2 \right ) \frac{Q^2}{v_0}, \nonumber \\
   V_{T'}&=&\frac{1}{\rho'}\left (1-\frac{\omega \rho'}{q}\delta^2 \right )\frac{Q^2}{v_0}, 
\end{eqnarray}
with the dimensionless factors 
\begin{equation}
\delta=\frac{m_\mu}{ \sqrt{|Q^2|}},
\kern 1cm
\rho=\frac{|Q^2|}{q^2}, 
\kern 1cm 
\rho'= \frac{q}{\epsilon+\epsilon'}. 
\end{equation}

The five nuclear response functions only depend on $(q,\omega)$, and
are the following components of the hadronic
tensor,
\begin{eqnarray}
  R_{CC}&=&W^{00},\nonumber \\  
  R_{CL}&=&-\frac{1}{2}(W^{03}+W^{30}), \nonumber \\
  R_{LL}&=&W^{33},\nonumber \\ 
  R_{T}&=&W^{11}+W^{22},\nonumber \\
  R_{T'}&=&-\frac{i}{2}(W^{12}-W^{21}).
\label{responses}
\end{eqnarray}
In the case of charged-current weak interactions, the nuclear current
operator is the sum of a vector and an axial-vector component. As a
result, the response functions CC, CL, LL, and T, 
can each
be written as the sum of two separate contributions: one arising from
the vector-vector (VV) part of the current, and the other from the
axial-axial (AA) part:
\begin{equation}
  R_K = R_K^{VV} + R_K^{AA}, \;\;\;\;\;\;\;\; K=CC,\, CL,\, LL,\, T \, .
\end{equation}
On the other hand, the $T'$ response originates from the interference
between the vector and axial components of the current. 
and can be written as
\begin{equation}
  R_{T'} = R_{T'}^{VA} + R_{T'}^{AV} \, .
\end{equation}

\subsection{Hadronic tensor}

We now describe the construction of the hadronic tensor in the 1p1h
sector within the RFG model, which provides a
transparent and analytically tractable framework for studying MEC
effects.
In this case, the hadronic tensor is
constructed from the matrix elements of the nuclear current between
the ground state of the RFG and excited 1p1h states, which are
represented by Slater determinants built from plane-wave
single-particle states within a volume $V$
\begin{eqnarray}
W^{\mu\nu}_{1p1h}&=& \sum_{ph}
\left\langle
ph^{-1} \right|\hat{J}^{\mu}(\nq) |\left. F \right\rangle^{*}
\left\langle
ph^{-1} \right|\hat{J}^{\nu}(\nq) |\left. F \right\rangle 
\nonumber
\\ 
&\times& \delta(E_{p}-E_{h}-\omega)
\theta(p-k_F)\theta(k_F-h).
\label{hadronic}
\end{eqnarray}
where $E_p=\sqrt{p^2+m_N^2}$ and $E_h=\sqrt{h^2+m_N^2}$ are the
on-shell energies of the nucleons involved in the 1p-1h excitation,
with momenta $\np$ and $\nh$, respectively.  The step functions
ensure that the initial nucleons have a momentum below the Fermi
momentum $h< k_F$  and the final nucleons have a momentum \( p > k_F
\).  The sums in Eq. (\ref{hadronic}) include sums over the spin
projections \( s_p \), \( s_h \) and isospin states \( t_p \), \( t_h
\) of the particle and hole, respectively. However, in the case of
charged-current neutrino scattering, the weak interaction selects
specific isospin transitions: the initial hole must be a neutron and
the final particle a proton
(the opposite holds for antineutrino scattering).

The nuclear current  is taken as the sum of 
one-body (1b) and  two-body (2b) operators.
\begin{equation}
\hat{J}^\mu = 
\hat{J}^\mu_{1b} 
+\hat{J}^\mu_{2b}.
\end{equation}
The 1p1h  matrix element of these operators in the RFG is given by 
\begin{equation}
\left\langle ph^{-1} \right|\hat{J}_{1b}^{\mu} |\left. F \right\rangle
=
\left\langle p \right|\hat{J}_{1b}^{\mu} |\left. h \right\rangle,
\label{j1b}
\end{equation}
\begin{equation}
\left\langle ph^{-1} \right|\hat{J}_{2b}^{\mu} |\left. F \right\rangle 
=
\sum_{k<k_F}\left[
\left\langle pk \right|\hat{J}_{2b}^{\mu} |\left. hk \right\rangle 
- \left\langle pk \right|\hat{J}_{2b}^{\mu} |\left. kh \right\rangle
\right].
\label{j2b}
\end{equation}
Note that the matrix element of the MEC, being a two-body operator,
involves a transition between pairs of nucleons; however, in a 1p1h
excitation, one of the nucleons,  $|k\rangle= |k,s_k,t_k\rangle $,
remains in its initial state and acts merely as a spectator.

The next step is to write the elementary matrix elements of the
one-body and two-body current operators between plane-wave states, using
momentum conservation 
 \begin{eqnarray}
 \langle p |\hat{J}_{1b}^{\mu} | h\rangle 
&=&
  \frac{(2\pi)^{3}}{V}\delta^{3}(\nq+\nh-\np)
j_{1b}^{\mu}(\np,\nh), 
\label{OBmatrix}
\\
\langle p'_{1}p'_{2}|\hat{J}_{2b}^{\mu}|p_{1}p_{2}\rangle
&=&
\frac{(2\pi)^{3}}{V^{2}}\delta^{3}(\np_1+\np_2+\nq-\np'_1-\np'_2)
\nonumber\\
&&
%\kern -1.5cm 
{}\times
j_{2b}^{\mu}(\np'_1,\np'_2,\np_1,\np_2) , 
\label{TBmatrix}
\end{eqnarray}
The current functions $j^\mu_{1b}(\np,\nh)$ and \(j_{2b}^{\mu}(\np'_1,
\np'_2, \np_1, \np_2)\) implicity depend on spin and isospin indices.
Inserting Eqs. (\ref{OBmatrix}) and (\ref{TBmatrix}) in
Eqs. (\ref{j1b}) and (\ref{j2b}), respectively, we can write the total
current matrix element in the form
 \begin{equation}
\left\langle ph^{-1} \right|\hat{J}^{\mu} |\left. F \right\rangle 
=
  \frac{(2\pi)^{3}}{V}\delta^{3}(\nq+\nh-\np)
j^{\mu}(\np,\nh), 
\label{total}
\end{equation}
 where
\begin{equation}
j^{\mu}(\np,\nh) \equiv j_{1b}^{\mu}(\np,\nh)+ j_{2b}^{\mu}(\np,\nh), 
\end{equation}
and
\begin{eqnarray}
j_{2b}^{\mu}(\np,\nh) 
\equiv 
\frac{1}{V}
\sum_{k<k_F}
\left[ j_{2b}^{\mu}(\np,\nk,\nh,\nk)-j_{2b}^{\mu}(\np,\nk,\nk,\nh)\right].
\label{effectiveOB}
\end{eqnarray}
This effective current, $j_{2b}^\mu(\np,\nh)$, accounts for the fact that the
two-body current operator, when acting on a nucleon pair in
the Fermi sea, can lead to a 1p1h excitation if one of the nucleons
remains a spectator. This mechanism is responsible for the
interference contribution in the 1p1h channel.
Then the total current function $j^\mu(\np,\nh)$
incorporates the contribution of
the 1b and the 2b currents.

To evaluate the hadronic tensor (\ref{hadronic}) in the RFG
we insert the matrix element (\ref{total}), 
and take the thermodynamic limit by replacing the discrete sum over hole
states with an integral over momentum space,
\begin{equation}
\sum_{h} \longrightarrow \frac{V}{(2\pi)^3} \int d^3h \sum_{s_h t_h}
\end{equation}
The integration over the
final particle states can then be performed using the
momentum-conserving delta function,
which fixes the particle momentum to $\mathbf{p} =
\mathbf{h} + \mathbf{q}$
\begin{eqnarray}
W^{\mu\nu}
&=&
\frac{V}{(2\pi)^{3}}
\int d^3h\delta(E_{p}-E_{h}-\omega)
2 w^{\mu\nu}(\np,\nh) \nonumber \\
&\times&
\theta(p-k_{F})\theta(k_{F}-h),
\label{integralw}
\end{eqnarray}  
where
\begin{equation}
w^{\mu\nu}(\np,\nh)\equiv\frac{1}{2}\sum_{s_ps_h}
j^\mu(\np,\nh)^*j^\nu(\np,\nh) 
\end{equation}
is the single-nucleon hadronic tensor.  The sums over isospin $t_p,t_h$ no
longer appear  because we have already imposed the condition
that, in the case of neutrino scattering, the hole state \( h \) corresponds
to a neutron and the particle \( p \) to a proton, while the opposite
holds for antineutrino scattering.

The single-nucleon tensor contains the square of the sum of the
one-body and two-body currents. By expanding this square, one obtains
\begin{equation}
w^{\mu\nu} = w^{\mu\nu}_{1b} + w^{\mu\nu}_{2b} + w^{\mu\nu}_{1b2b},
\end{equation}
where $w^{\mu\nu}_{1b}$ and $w^{\mu\nu}_{2b}$ are the pure one-body
and two-body contributions, respectively, and $w^{\mu\nu}_{1b2b}$ is
the interference term:
\begin{eqnarray}
w^{\mu\nu}_{1b}   & = & \frac{1}{2}\sum (j^\mu_{1b})^* j^\nu_{1b} \\
w^{\mu\nu}_{2b}   & = & \frac{1}{2}\sum (j^\mu_{2b})^* j^\nu_{2b} \\
w^{\mu\nu}_{1b2b} & = &
\frac12\sum [(j^\mu_{1b})^*j^\nu_{2b}+(j^\mu_{2b})^*j^\nu_{1b}].
\label{w1b2b}
\end{eqnarray}
The response functions also decompose accordingly:
\begin{equation}
R^K = R^K_{1b} + R^K_{2b} + R^K_{1b2b}
\end{equation}
From calculations in electron scattering, it has been observed that
the pure two-body MEC contribution, $R^K_{2b}$, is generally small and
can often be neglected when compared to the interference term
$R^K_{1b2b}$, which tends to dominate the MEC effects in the 1p1h
channel \cite{Ama03,Cas23}.  This justifies our focus on the
interference term, which captures the leading MEC effect
in the 1p1h sector.

The one-body current operator consists of two
terms: $j_{1b}^{\mu}(\np,\nh)=j_{1bV}^{\mu}(\np,\nh)+j_{1bA}^{\mu}(\np,\nh)$.
The vector current is
\begin{equation}
  j^{\mu}_{1bV}(\np,\nh)
=\bar{u}(\np)
\left(2F^V_{1}\gamma^{\mu}+i\frac{2F^V_{2}}{2m_{N}}\sigma^{\mu\nu}Q_{\nu}
\right)u(\nh),
\end{equation}
where the isovector nucleon form factors are defined as 
$F_{1,2}^V=(F_{1,2}^p-F_{1,2}^n)/2$.
The axial current is
\begin{equation}
  j^{\mu}_{1bA}(\np,\nh)
=-\bar{u}(\np)
\left(G_A\gamma^{\mu}\gamma_5 + G_P\frac{Q^\mu}{2m_{N}}\gamma_5
\right)u(\nh),
\label{axial}
\end{equation}
where $G_A$ is the nucleon axial-vector form factor and $G_P$ is the
pseudo-scalar form factor, given by
\begin{eqnarray}
  G_A&=&\frac{g_A}{(1-\frac{Q^2}{M_A^2})^2}
  \\
  G_P&=&\frac{4m_N^2}{m^2_\pi-Q^2}G_A
\end{eqnarray}
with $g_A=1.26$ and $M_A=1032$ MeV. 

Note that the minus sign in the axial one-body current (\ref{axial})
arises from our
convention of defining the total current as the sum of the vector and
axial parts, \( J^\mu = J_V^\mu + J_A^\mu \), whereas it is often
written in the literature as \( V - A \). We adopt this convention for
consistency with the meson-exchange currents, which are also defined
as the sum \( V + A \). Of course, physical results are independent of
this choice.

%====================================
\subsection{Meson exchange currents}
%-----------------------------------

\begin{figure}
\centering
\includegraphics[width=8cm,bb=120 310 495 700]{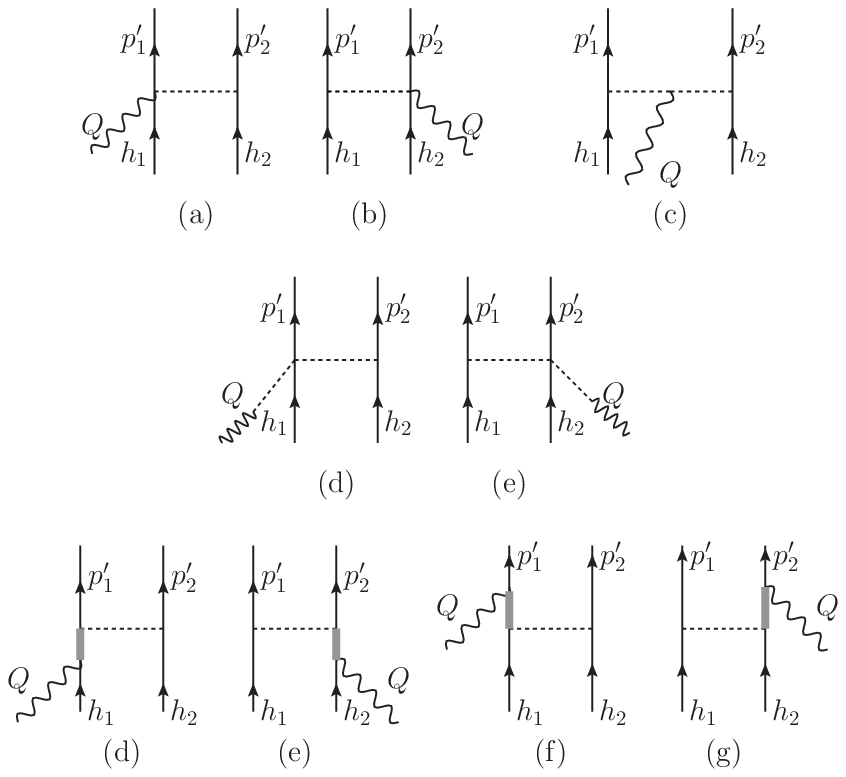}
\caption{Feynman diagrams for the electroweak MEC model used in 
this work.}\label{neudiag}
\end{figure}

In this section we present the MEC for CC neutrino scattering
considered in this work, corresponding to the Feynman diagrams shown
in Figure \ref{neudiag}.  They were derived in \cite{Rui17} from the
pion weak production model of ref. \cite{Her07}.  Similarly to the 1b
current, the weak MEC are the sum of vector and axial currents.  The
different current operators are: the seagull current (diagrams a and
b), the pion-in-flight current (diagram c), the pion-pole current
(diagrams d and e), and the $\Delta$ forward (f,g) and backward
currents (diagrams h-i):
\begin{equation}
j_2^\mu(\np'_1,\np'_2,\np_1,\np_2) 
=  j^{\mu}_{sea}+  j^{\mu}_{\pi}+  j^{\mu}_{pole}+  j^{\mu}_{\Delta}.
\end{equation}
All these currents share a
common structure involving the following spin matrix element
in each $\pi NN$ vertex: 
\begin{equation}
  V_{s'_1s_1}(p'_1,p_1) \equiv 
 F_{\pi NN}(k_{1}^{2})
\frac{\bar{u}_{s'_1}(\np'_1)\gamma^{5}\kbar_{1}u_{s_1}(\np_1)}{k_1^2-m_{\pi}^2},
\label{V-function}
\end{equation}
where $k_1= p'_1-p_1$, is the momentum transfer to the individual
nucleon (and the momentum carried by the pion),
the spinors \(u_s(\np)\) are the solutions of the Dirac equation with
momentum $\np$,
\( m_\pi \) is the pion mass,
and \( F_{\pi  NN}(k^2) \) is a strong form factor  \cite{Alb84,Som78,Mac87} 
\begin{equation}
  F_{\pi NN}(\mathbf{k}^2) = 
  \frac{\Lambda^{2}-m_{\pi}^{2}}{\Lambda^{2}-k^{2}},
  \kern 1cm \Lambda=1300\;\rm MeV
\end{equation}

\paragraph*{Seagull.}
%---------------------
The seagull current is given as the sum of the vector and axial operators,
\begin{eqnarray}
  j^{\mu}_{sea} &=& (j^{\mu}_{sea})_{V} + (j^{\mu}_{sea})_{A},  
\\
  (j^{\mu}_{sea})_V
&=&
i[\ntau^{(1)} \times \ntau^{(2)}]_{\pm}
\frac{f^2}{m_{\pi}^2}
F_{1}^{V} 
F_{\pi NN}(k_{1}^{2}) 
\nonumber\\
&&
\kern -1.5cm \times V_{s'_{1}s_{1}}(p'_{1},p_{1})
\bar{u}_{s'_2}(p'_{2}) \gamma^{5}\gamma^{\mu} u_{s_2}(p_{2}) 
 + (1 \leftrightarrow 2) 
\label{seaV}
\\
(j^{\mu}_{sea})_A
&=&
i[\ntau^{(1)} \times \ntau^{(2)}]_{\pm}
\frac{f}{m_{\pi}} \frac{F_{\rho}(k_{2}^{2})}{g_A}
F_{\pi NN}(k_{1}^{2}) 
\nonumber\\
&&\kern -1.5cm \times V_{s'_{1}s_{1}}(p'_{1},p_{1})
 \bar{u}_{s'_2}(p'_{2})\gamma^{\mu}u_{s_2}(p_{2}) 
 + (1 \leftrightarrow 2), 
\end{eqnarray}
where $f^2=1$ is the $\pi NN$ coupling constant, $\ntau^{(i)}$ is the
isospin operator of nucleon $i$, $F_1^V(Q^2) =F_1^p-F_1^n$ is the
isovector form factor of the nucleon, and $F_\rho$ is the $\rho$ meson
form factor. Note that the current has been written
as proportional to an isospin raising operator.
The isospin matrix elements are computed in Appendix A.

\paragraph*{Pion-in-flight.}
%--------------------------------
The pion-in-flight (or pionic) current has only a vector part
\begin{eqnarray}
(j^{\mu}_{\pi})_V &=&
i[\ntau^{(1)} \times \ntau^{(2)}]_{\pm}
F_{1}^{V}
\frac{f^2}{m_{\pi}^2}
V_{s'_{1}s_{1}}(p'_{1},p_{1})
\nonumber\\
&& \times
V_{s'_{2}s_{2}}(p'_{2},p_{2})(k_{1}^{\mu}-k_{2}^{\mu}) 
\\
(j^{\mu}_{\pi})_A&=&0  
\end{eqnarray}

\paragraph*{Pion-Pole.}
%-----------------------------
The pion-pole current is purely axial (this current could be
considered as the axial part of the pionic one)
\begin{eqnarray}
  (j^{\mu}_{pole})_V&=&0
  \\
  (j^{\mu}_{pole})_{A}&=&
i[\ntau^{(1)} \times \ntau^{(2)}]_{\pm}
\frac{f^2}{m_{\pi}^2}
\frac{F_{\rho}(k_{1}^{2})}{g_{A}}
F_{\pi NN}(k_{2}^{2})
\nonumber\\
&& \kern -1.5cm \times
\frac{Q^{\mu}\bar{u}_{s'_1}(p'_{1})\slashed{Q}u_{s_1}(p_{1})}{Q^2-m_{\pi}^2}
V_{s'_{2}s_{2}}(p'_{2},p_{2})
+ (1 \leftrightarrow 2)
\end{eqnarray}
Note that this current in proporcional to the four-momentum transfer, $Q^\mu$,
and it only contributes to the longitudinal and time components
of the hadronic tensor.

\paragraph*{Delta ($\Delta$).}
%-------------------------------------
The $\Delta$ excitation current operator has both
vector and axial parts,
\begin{equation}
 j^{\mu}_{\Delta}= (j^{\mu}_{\Delta})_V + (j^{\mu}_{\Delta})_A,
\end{equation}
corresponding to the vertices $\Gamma_V^{\beta\mu}$ and
$\Gamma_A^{\beta\mu}$, respectively,  in the 
\( N \rightarrow \Delta \) transition, given in  Eq. (\ref{tver}) below. 
The $\Delta$ current is further
divided into forward and backward operators
\begin{eqnarray}
  j^{\mu}_{\Delta F}
&=&
[U_{F}(1,2)_{\pm}]
\frac{f^{*}f}{m_{\pi}^2}
F_{\pi N \Delta}(k_{2}^{2})
V_{s'_{2}s_{2}}(p'_{2},p_{2})
\nonumber\\
&& \kern -1.5cm
\times
\bar{u}_{s'_1}(p'_{1})k_{2}^{\alpha}G_{\alpha\beta}(p_{1}+Q)
\Gamma^{\beta\mu}(Q)u_{s_1}(p_{1}) 
+ (1 \leftrightarrow 2), 
\label{deltaF}\\
  j^{\mu}_{\Delta B}
&=&
[U_{B}(1,2)_{\pm}]
\frac{f^{*}f}{m_{\pi}^2}
F_{\pi N \Delta}(k_{2}^{2})
V_{s'_{2}s_{2}}(p'_{2},p_{2})
\nonumber\\
&& \kern -1.5cm \times
\bar{u}_{s'_1}(p'_{1})k_{2}^{\beta}
\hat{\Gamma}^{\mu\alpha}(Q)G_{\alpha\beta}(p'_{1}-Q)u_{s_1}(p_{1}) 
+ (1 \leftrightarrow 2)
\label{deltaB}
\end{eqnarray}
where the $\pi N \Delta$
coupling constant is $f^*=2.13 f$.
The $\gamma N\Delta$ vertices are 
\begin{eqnarray}
  \Gamma^{\beta\mu}(Q)&=&
  \Gamma_V^{\beta\mu}(Q) + \Gamma_A^{\beta\mu}(Q),
\\
\Gamma_V^{\beta\mu}(Q) &=& \frac{C_3^V}{m_N} 
(g^{\beta\mu}\slashed{Q}-Q^{\beta}\gamma^{\mu})\gamma_5
 \label{tver}\\
\Gamma_A^{\beta\mu}(Q)&=& C_5^Ag^{\beta\mu}
\\
 \hat{ \Gamma}^{\beta\mu}(Q)&=&
  \hat{\Gamma}_V^{\beta\mu}(Q) + \hat{\Gamma}_A^{\beta\mu}(Q)
\\
\hat{\Gamma}_V^{\beta\mu}(Q) &=& -\Gamma_V^{\mu\beta}(Q),
\kern 1cm
\hat{\Gamma}_A^{\beta\mu}(Q) = \Gamma_A^{\beta\mu}(Q). 
\end{eqnarray}
The vector and axial form factors are taken from \cite{Her07}:
\begin{eqnarray}
  C_{3}^{V}(Q^{2})
&=&\frac{2.13}{(1-\frac{Q^{2}}{M_{V}^{2}})^{2}}
\frac{1}{1-\frac{Q^{2}}{4M_{V}^{2}}},\\
C_{5}^{A}(Q^{2})
&=&\frac{1.2}{(1-\frac{Q^{2}}{M_{A\Delta}^{2}})^{2}}
\frac{1}{1-\frac{Q^{2}}{4M_{A\Delta}^{2}}},
\end{eqnarray}
with $M_V=0.84$ GeV and $M_{A \Delta}=1.05$ GeV.
The $\pi$N$\Delta$ form factor
is taken as \cite{Alb84,Som78} 
\begin{equation}
  F_{\pi NN}(k)=   F_{\pi N\Delta}(k)
\label{pinnff}
\end{equation}

The forward $\Delta$ current corresponds to processes where the
$\Delta$ resonance is produced and then decays back to a nucleon,
while the backward $\Delta$ current involves the exchange of a pion,
leading to the creation of a $\Delta$ resonance in the intermediate
state. The charge dependence of these processes is embedded in the
isospin operators
$U_F(1,2)_{\pm} = U_{F}(1,2)_{x} \pm iU_{F}(1,2)_{y}$ for the forward term and
\( U_{B}(1,2)_{\pm} = U_{B}(1,2)_{x} \pm iU_{B}(1,2)_{y} \) for the backward
term, where
\begin{eqnarray}
 U_{F}(1,2)_i 
&=&
\sqrt{\frac{3}{2}}
\sum_{j=1}^{3}
T_{j}^{(1)}T_{i}^{(1)\dagger}\tau_{j}^{(2)},
\label{uf}\\ 
 U_{B}(1,2)_i 
&=&
\sqrt{\frac{3}{2}}
\sum_{j=1}^{3}
T_{i}^{(1)}T_{j}^{(1)\dagger}\tau_{j}^{(2)}.
\label{ub}
\end{eqnarray}
The operator $T_i^\dagger$ is the isospin raising operator that
connects isospin-$1/2$ states to isospin-$3/2$ states and satisfies
the condition
$T_i T_j^\dagger = \frac{2}{3} \delta_{ij} - \frac{i}{3} \epsilon_{ijk} \tau_k.$
Using this property it can be written
\begin{eqnarray}
  U_F(1,2)_\pm  &=&
   \frac{2}{\sqrt{6}} \tau_\pm^{(2)}
  -\frac{i}{\sqrt{6}} [\ntau^{(1)} \times \ntau^{(2)}]_{\pm}
\label{UF}  \\
    U_B(1,2)_\pm  &=&
   \frac{2}{\sqrt{6}} \tau_\pm^{(2)}
  +\frac{i}{\sqrt{6}} [\ntau^{(1)} \times \ntau^{(2)}]_{\pm}.
\label{UB}
\end{eqnarray}
Finally, the $\Delta$ propagator is
\begin{equation}
  G_{\alpha\beta}(P)=
\frac{{\cal P}_{\alpha\beta}(P)}{
P^{2}-m_\Delta^2+im_\Delta\Gamma(P^{2})+\frac{\Gamma(P^{2})^{2}}{4}}
\end{equation} 
where $m_\Delta$ and $\Gamma$ are the $\Delta$ mass and width
respectively. The projector ${\cal P}_{\alpha\beta}(P)$ over spin-3/2  is
\begin{eqnarray}
{\cal  P}_{\alpha\beta}(P)
&=& -(\Pbar+M_{\Delta})
\nonumber\\
&& \kern -1.5cm \times
\left[
  g_{\alpha\beta}-\frac{\gamma_{\alpha}\gamma_{\beta}}{3}
  -\frac{2P_{\alpha}P_{\beta}}{3m_\Delta^2}
  +\frac{P_{\alpha}\gamma_{\beta}-P_{\beta}\gamma_{\alpha}}{3m_\Delta}
\right] .
\end{eqnarray}  
The 
$\Delta$ width $\Gamma(P^{2})$ is given by
\begin{equation}
  \Gamma(P^{2})=\Gamma_{0}\frac{m_{\Delta}}{\sqrt{P^{2}}}
\left(\frac{p_{\pi}}{p^{res}_{\pi}}\right)^{3}.
\label{width}
\end{equation}
where $\Gamma_{0}=120$ MeV is the $\Delta$ width at rest, $p_{\pi}$ is
the momentum of the final pion in the $\Delta$ decay, and
$p^{res}_{\pi}$ is its value at resonance ($P^2=m_\Delta^2$).

The sign $\pm$ in the isospin matrix elements of
Eqs. (\ref{seaV}--\ref{deltaB}) refers to neutrino $(+)$ or
antineutrino $(-)$ scattering.

The vector part of the weak meson-exchange currents reduces to the
electromagnetic MEC when the isospin-raising operators are replaced by
their third components \cite{Cas23}.

\subsection{Calculation of the responses in the RFG}
%====================================================

In this section we outline the numerical procedure used to compute the
response functions in the relativistic Fermi gas model, including both
one-body and meson exchange currents.
First we perform
a change of variables to $(E_h, E_p, \phi)$ in Eq. (\ref{integralw}),
where $E_h$ and $E_p$ are
the energies of the hole and particle states, and $\phi$ is the
azimuthal angle. Then the volume element in spherical coordinates becomes
$h^2 dh d\cos\theta d\phi=
(E_hE_p/q)dE_hdE_p d\phi$.
The integration over $E_p$ can then be carried out
using the energy-conserving delta function, which fixes the polar
angle $\theta$
  between
  $\nq$ and $\nh$
\begin{equation} \label{angulo}
\cos\theta= \frac{2E_h\omega+Q^2}{2hq}. 
\end{equation}
Since the \( z \)-axis is chosen along the direction of the momentum
transfer \( \mathbf{q} \), the integrand is independent of the
azimuthal angle \( \phi \). We choose $\phi=0$ and the integration
over \( \phi \) yields a factor of \( 2\pi \). The remaining integral
is then performed numerically over the hole energy \( E_h \).
 \begin{equation}
R_K(q,\omega)
= 
 \frac{V}{(2 \pi)^3}
 \frac{2\pi m_N^3}{q}
\int_{\epsilon_0}^{\infty}d\epsilon\, n(\epsilon)\, 2w_K(\epsilon,q,\omega),
\label{respuesta}
\end{equation}
where we have defined the adimensional energies $\epsilon=E_h/m_N$ and
$\epsilon_F=E_F/m_N$.  Moreover we have introduced the energy
distribution of the Fermi gas $n(\epsilon)=
\theta(\epsilon_F-\epsilon)$.  The lower limit of the integral
(\ref{respuesta}), denoted \( \epsilon_0 \), corresponds to the
minimum energy of an on-shell nucleon that can absorb the energy and
momentum transfer \( (\omega, \mathbf{q}) \). It is given by
\cite{Ama20}
\begin{equation} \label{epsilon0}
\epsilon_0={\rm Max}
\left\{ 
       \kappa\sqrt{1+\frac{1}{\tau}}-\lambda, \epsilon_F-2\lambda
\right\},
\end{equation}
where we use the dimensionless variables 
\begin{equation}
\lambda  = \frac{\omega}{2m_N},
\kern 1cm 
\kappa   =  \frac{q}{2m_N}, 
\kern 1cm
\tau  =  \kappa^2-\lambda^2. 
\end{equation}
It can be seen that the minimum energy \( \epsilon_0 \) is always
greater than or equal to one, which corresponds to a nucleon at
rest. This situation defines the center of the QE peak,
$\omega=Q^2/(2m_N)$. It is also easy to verify that this occurs when
\( \lambda = \tau \), in which case \( \kappa^2 = \tau(\tau + 1) \).

The integrand in Eq.~(\ref{respuesta}) is the single-nucleon response
\( w_K(\epsilon,q,\omega) \) corresponding to a 1p1h excitation,
obtained from the hadronic tensor of a single nucleon using the same
definition as in Eqs.~(\ref{responses}). This function depends in
general on the momenta \( \mathbf{h} \) and \( \mathbf{p} \), but in
practice it only depends on three variables. Once \( \epsilon \) is
fixed, the hole energy is given by \( E_h = m_N \epsilon \), from
which the magnitude of the momentum \( h = |\mathbf{h}| \) can be
determined. For fixed \( q \) and \( \omega \), the scattering angle
\( \cos\theta \) can then be calculated using (\ref{angulo}),
which determines the
direction of \( \mathbf{h} \) (we take \( \phi = 0 \) without loss of
generality). By adding the momentum transfer \( \mathbf{q} \), the
final momentum \( \mathbf{p} = \mathbf{h} + \mathbf{q} \) is fully
specified.

In the RFG model, the total current \( j^\mu(\mathbf{p}, \mathbf{h})
\), including the 1b and 2b currents, and the single-nucleon responses
$w_K$ are computed numerically. This requires performing a numerical
integration over the three-momentum \( \mathbf{k} \) of the
intermediate nucleon, as well as sums over the spin projections \( s_k
\), \( s_p \), and \( s_h \).

\begin{figure}
\centering
\includegraphics[width=8cm,bb=110 310 500 690]{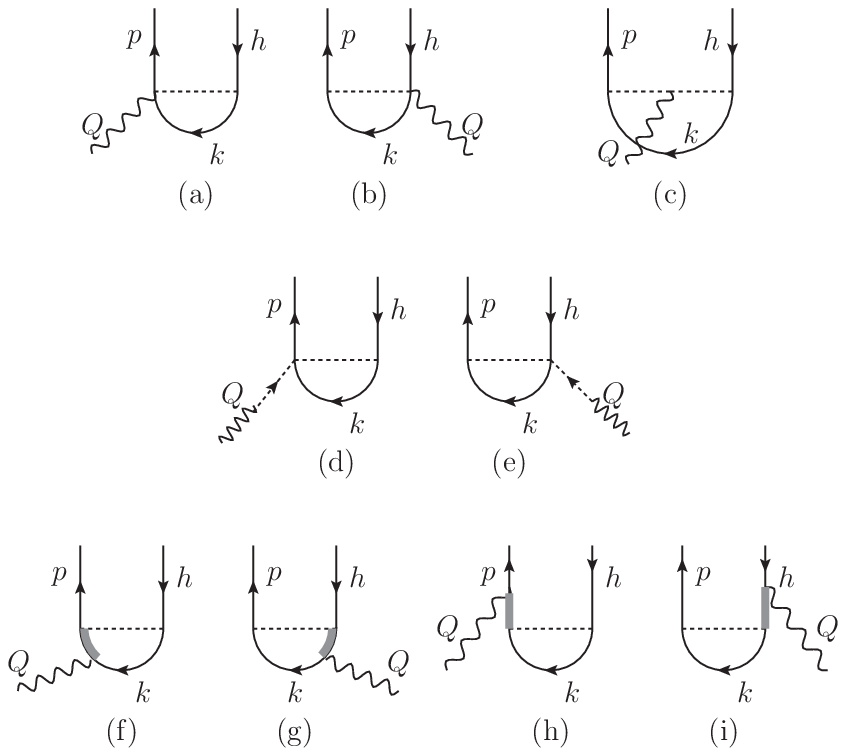}
\caption{Exchange many-body diagrams of the MEC matrix elements
  contributing to the 1p1h excitation channel considered in this
  work.}
\label{feyneu}
\end{figure}

The isospin sums over \( t_k \) that appear in the two-body current
are provided in Appendix~\ref{appendix:isospin}.  The resulting
effective one-body current
consists of a direct term minus an exchange term
\begin{equation}
j_{2b}^{\mu}(\np,\nh) =
j_{2b}^{\mu}(\np,\nh)_{\rm dir}- 
j_{2b}^{\mu}(\np,\nh)_{exch} 
\end{equation}
The exchange contribution corresponds to the diagrams
shown in Fig.~\ref{feyneu}. In symmetric nuclear matter, the
direct term of the vector current vanishes---similarly to the case of
the electromagnetic exchange current---because it arises solely from the
\(\Delta\) current, which is transverse, and therefore vanishes upon
contraction with \( Q^\mu \). When summing over spin in the direct
matrix element, this contraction appears explicitly, leading to a
vanishing contribution.

However, for the axial \(\Delta\) current, the direct term does not
vanish, and in principle should be included when computing the
neutrino response. Nevertheless, we have checked that this term
contributes significantly only to the longitudinal response \( R_{LL}
\), which is known to be small. This behavior can be seen explicitly
in the non-relativistic limit, where the structure of the current
becomes more transparent.  In the results section we show through
explicit calculations that the LL response
gives a negligible contribution to the cross section for quasielastic
neutrino scattering. Since the direct axial MEC term contributes
significantly only to the LL channel, and this channel plays a minor
role in the kinematics of interest, it is not included in the present
work.

%-----------------------------------
\section{Non-relativistic limit}
%-----------------------------------

In this section we present the expressions for the interference responses in the non-relativistic (NR) limit, which will be compared with the corresponding relativistic results in Section IV.
This comparison serves several purposes: (i) to verify
that relativistic and non-relativistic results agree at low momentum
and energy transfer, as expected; (ii) to compare our results with
known expressions for axial MEC currents in the NR literature
\cite{Ris89}; and (iii) to test the accuracy of our numerical
implementation, since the NR Fermi gas allows partial analytical
evaluation of the response functions
\cite{Ama94a,Ama94b}. Furthermore, this limit helps identify the
dominant responses and current components at moderate momentum
transfer.

\subsection{Non-relativistic one-body current}
%===================================================================

The vector part of the 1b current is the sum of magnetization and convection
currents:
\begin{eqnarray}
  \nj_{1b}(\np,\nh)_V
&=& 
  \nj_{mV}(\np,\nh)+  \nj_{cV}(\np,\nh), 
\\
  \nj_{mV}(\np,\nh)
&=&
-\frac{2G_M^V}{2m_N}i\nq\times\nsigma_{s_ps_h},
\label{magnetization}
\\
  \nj_{cV}(\np,\nh)
&=&
\delta_{s_ps_h}
\frac{2G_E^V}{m_N}(\nh+\frac{\nq}{2}),
\label{convection}
\end{eqnarray}
with $\nq=\np-\nh$ by momentum conservation.  Here $G_M^V$ ($G_E^V$)
is the isovector magnetic (electric) form factor of the nucleon,
$G_E^V=(G_E^p-G_E^n)/2.$ 

The transverse (perpendicular to $\nq$) 
one-body axial current at leading order is \cite{Ama05}
\begin{equation}
\nj^\perp_{1b}(\np,\nh)_{A}=-G_A \nsigma^\perp_{s_ps_h}.
\end{equation}

The time component of the axial current, $j^0_{1b}$,
or axial charge-density is typically not included in non-relativistic
calculations because it is small at leading order; however, this
suppression is not a strict consequence of the non-relativistic
expansion itself.  To obtain its static
limit, we have considered the semi-relativistic expansion of the
electroweak current introduced in
Ref.~\cite{Ama96,Ama05}.
It can be written
as the sum of a convective term and a magnetization term, in
analogy with the structure of the transverse vector current \cite{Ama05}
\begin{eqnarray}
  j^0_{1b}(\np,\nh)_{A} & = & j^0_{mA}(\np,\nh)+j^0_{cA}(\np,\nh) \\
 j^0_{mA}(\np,\nh)     & = & - \frac{G_A'}{2m_N}\nq \cdot \nsigma_{s_ps_h}\\
 j^0_{cA}(\np,\nh)     & = &
-G_A\frac{\nh^\perp}{m_N} \cdot \nsigma_{s_ps_h}
\label{obaxial}
\end{eqnarray}
where $\nh^\perp$ is the transverse component of $\nh$ ---perpendicular
to the momentum transfer. The auxiliar form factor \( G_A' \) is
defined by
\begin{equation}
  G_A'= G_A - \tau G_P=\left( 1- \frac{Q^2}{Q^2-m_\pi^2}\right)G_A.
\end{equation}

\subsection{Non-relativistic expansion of the weak MEC}
%=================================-----------------------------------

The weak MEC operators are the sum of
vector plus axial components. The vector MEC are also isovectors
and closely related to the electromagnetic MEC, which are also
isovectors. The key difference lies in the isospin structure: while
the electromagnetic current corresponds to the third component
of the isospin operator, the charged weak current
involves the raising and lowering \( \pm \) operators.
Consequently, the non-relativistic expressions for electromagnetic MEC
derived in Ref. \cite{Cas25} can be directly adapted to the weak case by
substituting  \( \tau_z  \rightarrow \tau_\pm \), and
$[\ntau^{(1)} \times \ntau^{(2)}]_z\rightarrow [\ntau^{(1)} \times \ntau^{(2)}]_\pm$,
depending on the specific charge-changing process.

%    from paper 3
The non-relativistic expansion of the MEC is obtained by applying
standard reduction rules to matrix elements involving products of
gamma matrices between Dirac spinors, retaining only leading-order
terms in $1/m_N$:
\begin{eqnarray}
\gamma^0 \longrightarrow 1, &
\gamma^i \longrightarrow 0, &
\gamma_5\gamma^0 \longrightarrow 0, 
\label{gammas1}\\
\gamma_5\gamma^i \longrightarrow -\sigma_i, &
\gamma^i\gamma^j \longrightarrow -\sigma_i\sigma_j, &
\gamma^0\gamma^j \longrightarrow 0. 
\label{gammas2}
\end{eqnarray}
For a nucleon momentum: 
\begin{eqnarray}
 p^\mu \longrightarrow (m_N,p^i), &&
\pbar \longrightarrow p_0. 
\end{eqnarray}
For the momentum transfer to nucleon $i$:
\begin{eqnarray}
k^\mu \longrightarrow (0,k^i), &&
\gamma_5\kbar \longrightarrow \nk\cdot\nsigma.
\end{eqnarray}
Finally, the $V$-function of Eq. (\ref{V-function}) becomes:
\begin{equation}
  V(1',1) \longrightarrow 
-\frac{\nk_1\cdot\nsigma^{(1)}}{\nk_1^2+m_{\pi}^2}.
\label{Vnorel}
\end{equation}

As a result, only the spatial components of the vector MEC survive at
leading order in the non-relativistic expansion. This feature is also
confirmed with the fully relativistic calculation. Then 
\begin{eqnarray}
j^{\mu}_{s V} 
&\overset{\text{nr}}{\longrightarrow}& 
(0,\nj_{s V}) \nonumber
\\
j^{\mu}_{\pi V} 
&\overset{\text{nr}}{\longrightarrow}& 
(0,\nj_{\pi V}) \nonumber
\\
j^{\mu}_{\Delta V} 
&\overset{\text{nr}}{\longrightarrow}& 
(0,\nj_{\Delta V}) \nonumber
\end{eqnarray}

The axial seagull current is proportional to the matrix element of \(
\gamma^\mu \), and in the non-relativistic limit, only its time
component (proportional to \( \gamma^0 \)) survives.  On the other
hand, only the spatial components of the axial \(\Delta\) current
remain non-zero, as shown in appendix B.  The pion-pole current,
being proportional to $q_\mu \gamma^\mu$,
vanishes at leading order.
The surviving axial currents are
\begin{eqnarray}
j^{\mu}_{sA} 
&\overset{\text{nr}}{\longrightarrow}& 
(j^{0}_{sA},\vec{0}) \nonumber 
\\
j^{\mu}_{\Delta A} 
&\overset{\text{nr}}{\longrightarrow}& 
(0,\nj_{\Delta A}) \nonumber
\end{eqnarray}
The corresponding non-relativistic operators are

\begin{widetext}

\begin{eqnarray}
  \nj_{sV}(p'_1,p'_2,p_1,p_2)
  &=& 
i[\ntau^{(1)} \times \ntau^{(2)}]_+
\frac{f^{2}}{m_{\pi}^{2}}F_1^V
\left(
\frac{\nk_1\cdot\nsigma^{(1)}}{\nk_1^2+m_{\pi}^2}
\nsigma^{(2)}
-\frac{\nk_2\cdot\nsigma^{(2)}}{\nk_2^2+m_{\pi}^2}
\nsigma^{(1)}
\right)
\label{seagull}
\\
\nj_{\pi V}(p'_1,p'_2,p_1,p_2) &=&
  i[\ntau^{(1)} \times \ntau^{(2)}]_+
\frac{f^{2}}{m_{\pi}^{2}}F_1^V
\frac{\nk_1\cdot\nsigma^{(1)}}{\nk_1^2+m_{\pi}^2}
\frac{\nk_2\cdot\nsigma^{(2)}}{\nk_2^2+m_{\pi}^2}
(\nk_1-\nk_2)
\label{pionic}
\\
 \nj_{\Delta V}(p'_1,p'_2,p_1,p_2) &=&
i \sqrt{ \frac32 } \frac29  \frac{ff^*}{m_\pi^2}
\frac{C_3^V}{m_N}\frac{1}{m_\Delta-m_N}
\left\{
\frac{\nk_2\cdot\nsigma^{(2)}}{\nk_2^2+m_{\pi}^2}
\left[
4\tau_{+}^{(2)}\nk_2+
[\ntau^{(1)}\times\ntau^{(2)}]_{+}
\nk_2\times\nsigma^{(1)}
\right]
\right.
\nonumber\\
&&
\kern 4cm 
\left. \mbox{}+
\frac{\nk_1\cdot\nsigma^{(1)}}{\nk_1^2+m_{\pi}^2}
\left[
4\tau_{+}^{(1)}\nk_1-
[\ntau^{(1)}\times\ntau^{(2)}]_{+}
\nk_1\times\nsigma^{(2)}
\right]
\right\}
\times\nq
\label{deltafinal}
\\
  j^{0}_{sA}(p'_1,p'_2,p_1,p_2) &=&
  -i[\ntau^{(1)} \times \ntau^{(2)}]_+
  \frac{f^{2}_{\pi  NN}}{m_{\pi}^{2}}\frac{1}{g_A}
  \left[ \frac{ \nk_{1}\cdot\nsigma^{(1)} }{ \nk_{1}^{2}+m^{2}_{\pi} }
        -\frac{ \nk_{2}\cdot\nsigma^{(2)} }{ \nk_{2}^{2}+m^{2}_{\pi} }
  \right]
 \\
\nj_{\Delta A}(p'_1,p'_2,p_1,p_2) &=&
-\sqrt{\frac{3}{2}}\frac{2}{9}\frac{ff^{*}}{m_{\pi}^{2}}C_{5}^{A}
\frac{1}{m_\Delta-m_N}
\Biggl\{
4\tau_{+}^{(1)}\frac{ ( \nk_{1} \cdot \nsigma^{(1)} )\nk_{1} }{ \nk_1^2+m^{2}_\pi }
+ 4\tau_{+}^{(2)}\frac{ ( \nk_{2} \cdot \nsigma^{(2)} )\nk_{2} }{ \nk_2^2+m^{2}_\pi }
  \nonumber \\
  &&
  +[\ntau^{(1)}\times\ntau^{(2)}]_{+}
  \biggl[
  \frac{(\nk_{2}\cdot\nsigma^{(2)})(\nk_{2}\times\nsigma^{(1)})}{\nk_2^2+m^{2}_\pi}
  -\frac{(\nk_{1}\cdot\nsigma^{(1)})(\nk_{1}\times\nsigma^{(2)})}{\nk_1^2+m^{2}_\pi}
  \biggl]
  \Biggl\}. 
\label{deltaAxial}
\end{eqnarray}
These operators match the standard non-relativistic MEC in the
literature \cite{Ris89} modulo differences in coupling constants and
form factors.

\subsection{MEC effective one-body currents}

The 1p1h matrix elements of the vector and axial MEC are
\begin{equation}
j^\mu_{2b}(p,h)=
-\int \frac{d^3k}{(2\pi)^3}
\sum_{t_ks_k}j^\mu_{2b}(p,k,k,h)
=
j^\mu_{s}(p,h)+j^\mu_{\pi}(p,h)+j^\mu_{\Delta}(p,h),
\end{equation}
where we have neglected the direct part in the axial $\Delta$ current,
as previously mentioned. After preforming the sums over $t_k, s_k$,
the results are the following for the three MEC, seagull, pionic and
$\Delta$ currents
\begin{eqnarray}
\nj_s(p,h)_V
&=&
4
\frac{f^2}{m_\pi^2}F_1^V
\int \frac{d^3k}{(2\pi)^3}
\left(
\frac{\delta_{s_ps_h}\nk_1+i\nsigma_{ph}\times\nk_1}{\nk_1^2+m_{\pi}^2}
-\frac{\delta_{s_ps_h}\nk_2+i\nk_2\times\nsigma_{ph}}{\nk_2^2+m_{\pi}^2}
\right)
\label{seagullph}
\\
\nj_\pi(p,h)_V
&=&
4
\frac{f^2}{m_\pi^2}F_1^V
\int \frac{d^3k}{(2\pi)^3}
\frac{\delta_{s_ps_h}\nk_1\cdot\nk_2
+i(\nk_1\times\nk_2)\cdot\nsigma_{ph}}
{(\nk_1^2+m_{\pi}^2)(\nk_1^2+m_{\pi}^2)}(\nk_1-\nk_2),
\label{pionicph}
\\
\nj_\Delta(p,h)_V
&=&
i
 \sqrt{ \frac32 } \frac89  \frac{ff^*}{m_\pi^2}
\frac{C_3^V}{m_N}\frac{1}{m_\Delta-m_N}
\nq\times 
\int \frac{d^3k}{(2\pi)^3}
\left(
\frac{\nk_1^2\nsigma_{ph}+(\nsigma_{ph}\cdot\nk_1)\nk_1}
     {\nk_1^2+m_\pi^2}
+\frac{\nk_2^2\nsigma_{ph}+(\nsigma_{ph}\cdot\nk_2)\nk_2}
     {\nk_2^2+m_\pi^2}
\right),
\label{deltaph}
\\
  j_{s}^0(p,h)_A 
&=&
-4\frac{f^{2}}{m_{\pi}^{2}}\frac{1}{g_A}
\int \frac{d^3k}{(2\pi)^3}
\left[ 
     \frac{ \nk_{1} \cdot \nsigma_{ph}  }{  \nk_{1}^{2}+m^{2}_{\pi} }
    -\frac{ \nk_{2} \cdot \nsigma_{ph}  }{  \nk_{2}^{2}+m^{2}_{\pi} } 
\right] 
\\
\nj_{\Delta}(p,h)_A 
&=& 
 \sqrt{ \frac32 } \frac89  \frac{ff^*}{m_\pi^2}
C_5^A\frac{1}{m_\Delta-m_N}
\int \frac{d^3k}{(2\pi)^3}
\left(
\frac{\nk_1^2\nsigma_{ph}+(\nsigma_{ph}\cdot\nk_1)\nk_1}
     {\nk_1^2+m_\pi^2}
+\frac{\nk_2^2\nsigma_{ph}+(\nsigma_{ph}\cdot\nk_2)\nk_2}
     {\nk_2^2+m_\pi^2}
\right),
\end{eqnarray}
with $\nk_1=\np-\nk$,  $\nk_2=\nk-\nh$ and $\nsigma_{ph}=\nsigma{s_ps_h}$.

\end{widetext}

\subsection{Non relativistic 1b2b interference single-nucleon responses}
%----------------------------------------------------
Here we provide separate expressions for the
different contributions to the interference terms, which arise from
the cross products between the various components of the one-body
current and the meson exchange currents.
\begin{eqnarray}
w_{CC,1b2b}^{AA}&=& w_{CC,ms}^{AA}+ w_{CC,cs}^{AA}\\ 
\nonumber\\
  w_{CC,ms}^{AA}  &=&  
{\rm Re}\sum_{s_ps_h}
j^0_{mA}(p,h)^*\; j_{sA}^0(p,h) \\
   w_{CC,cs}^{AA} 
 &=&  {\rm Re}\sum_{s_ps_h}j^0_{cA}(p,h)^* \; j_{sA}^0(p,h)
\\
w_{T,1b2b}^{VV}&=& 
w_{T,ms}^{VV} +
w_{T,cs}^{VV} +
w_{T,m\pi}^{VV}+
w_{T,c\pi}^{VV}+ 
w_{T,m\Delta}^{VV}
\nonumber \\
\end{eqnarray}
\begin{eqnarray}
w_{T,ms}^{VV} &=& {\rm Re}\sum_{s_ps_h} \nj_{mV}(p,h)^*\cdot\nj_{sV}(p,h)
\\
w_{T,cs}^{VV} & = & {\rm Re}\sum_{s_ps_h} \nj^T_{cV}(p,h)^*\cdot\nj_{sV}(p,h) 
\\
w_{T,m\pi}^{VV}&  = & {\rm Re}\sum_{s_ps_h} \nj_{mV}(p,h)^*\cdot\nj_\pi(p,h) 
\\
w_{T,c\pi}^{VV} & = & {\rm Re}\sum_{s_ps_h} \nj^T_{cV}(p,h)^*\cdot\nj_\pi(p,h) 
\\
w_{T,m\Delta}^{VV}&  = & {\rm Re}\sum_{s_ps_h} \nj_{mV}(p,h)^*\cdot\nj_{\Delta V}(p,h) 
\\
w_{T,1b2b}^{AA} & =&   w_{T,1b\Delta}^{AA} = 
 {\rm Re}\sum_{s_ps_h}\nj^\perp_{1b A}(p,h)^*\cdot \nj_{\Delta A}(p,h) 
\nonumber\\
\end{eqnarray}

In the case of the $T'$ response we separate the VA and AV contributions
\begin{eqnarray}
  w_{T',1b2b} &=& 
  w_{T',1b2b}^{VA}+ 
  w_{T',1b2b}^{AV}
\\
  w_{T',1b2b}^{VA}
& = &
\frac12 \mbox{Im} \sum_{s_ps_h}
\Big[  j_{1bV}^{1*}(p,h) j_{2bA}^{2}(p,h)
\nonumber\\
&&
+  j_{2bA}^{1*}(p,h) j_{1bV}^{2}(p,h) \Big]
\\
  w_{T',1b2b}^{AV}
& = &
\frac12 \mbox{Im} \sum_{s_ps_h}
\Big[  j_{1bA}^{1*}(p,h) j_{2bV}^{2}(p,h)
\nonumber\\
&&
+  j_{2bV}^{1*}(p,h) j_{1bA}^{2}(p,h) \Big]
\end{eqnarray}
Then
\begin{eqnarray}
  w_{T',1b2b}^{VA} &=&
  w_{T',m\Delta}^{VA}
\\
  w_{T',1b2b}^{AV} &=&
  w_{T',1b s}^{AV}+
  w_{T',1b \pi}^{AV}+
  w_{T',1b \Delta}^{AV}
\end{eqnarray}
Note that in the case of the $\Delta$ current, both the vector and
axial parts contribute to the response $T'$, whereas in the case of the
seagull and pionic current, only the vector part is considered, since the axial
part is longitudinal in the non-relativistic limit.  

The explicit expressions for the different contributions to the single
nucleon interference responses are the following
(with  $\nk_1=\np-\nk$ and $\nk_2=\nk-\nh$):
\begin{widetext}

\begin{eqnarray}
w^{VV}_{T,ms}
&=&
2\frac{f^2}{m_\pi^2}F_1^V\frac{G_M^V}{m_N}
\int \frac{d^3k}{(2\pi)^3}
\left(
\frac{4\nq\cdot\nk_1}{\nk_1^2+m_{\pi}^2}+
\frac{4\nq\cdot\nk_2}{\nk_2^2+m_{\pi}^2}
\right),
\label{wms}
\\
w^{VV}_{T,cs}
&=&
4\frac{f^2}{m_\pi^2}F_1^V\frac{G_E^V}{m_N}
\int \frac{d^3k}{(2\pi)^3}
\left(
\frac{2\nh_T\cdot\nk_1}{\nk_1^2+m_{\pi}^2}-
\frac{2\nh_T\cdot\nk_2}{\nk_2^2+m_{\pi}^2}
\right),
\\
w^{VV}_{T,m\pi}
&=&
-2\frac{f^2}{m_\pi^2}F_1^V\frac{G_M^V}{m_N}
\int \frac{d^3k}{(2\pi)^3}
\frac{4(\nq\times\nk_2)^2}{(\nk_1^2+m_{\pi}^2)(\nk_2^2+m_{\pi}^2)},
\label{VVTmpi}
\\
w^{VV}_{T,c\pi}
&=&
-4\frac{f^2}{m_\pi^2}F_1^V\frac{G_E^V}{m_N}
\int \frac{d^3k}{(2\pi)^3}
\frac{4(\nq\cdot\nk_2-\nk_2^2)\nh_T\cdot\nk_2}{(\nk_1^2+m_{\pi}^2)(\nk_2^2+m_{\pi}^2)},
\\
w^{VV}_{T,m\Delta}
&=&
-2  \sqrt{ \frac32 } \frac29  \frac{ff^*}{m_\pi^2}
\frac{C_3^V}{m_N}\frac{1}{m_\Delta-m_N} \frac{G_M^V}{m_N}
\int \frac{d^3k}{(2\pi)^3}
2\left(
\frac{3q^2k_1^2-(\nq\cdot\nk_1)^2}{\nk_1^2+m_{\pi}^2}+
\frac{3q^2k_2^2-(\nq\cdot\nk_2)^2}{\nk_2^2+m_{\pi}^2}
\right),
\label{wdelta1}
\\
w_{CC,ms}^{AA} 
&=&
\frac{f^2}{m_{\pi}^{2}}\frac{1}{g_A}\frac{G_A'}{m_N}
\int \frac{d^3k}{(2\pi)^3}
\left(
\frac{4\nq\cdot\nk_1}{\nk_1^2+m_{\pi}^2}+
\frac{4\nq\cdot\nk_2}{\nk_2^2+m_{\pi}^2}
\right),
\\
w_{CC,cs}^{AA}
&=&
4 \frac{f^2}{m_{\pi}^{2}}\frac{1}{g_A}
\frac{G_A}{m_N}
\int \frac{d^3k}{(2\pi)^3}
\left(
\frac{2\nh_T\cdot\nk_1}{\nk_1^2+m_{\pi}^2}-
\frac{2\nh_T\cdot\nk_2}{\nk_2^2+m_{\pi}^2}
\right),
\\
w_{T,1b\Delta}^{AA}
&=&
-\sqrt{ \frac32 } \frac{16}9  \frac{ff^*}{m_\pi^2}
\frac1{m_\Delta-m_N}
\frac{C_5^A G_A}{q^2}
\int \frac{d^3k}{(2\pi)^3}
2\left(
\frac{3q^2k_1^2-(\nq\cdot\nk_1)^2}{\nk_1^2+m_{\pi}^2}+
\frac{3q^2k_2^2-(\nq\cdot\nk_2)^2}{\nk_2^2+m_{\pi}^2}
\right),
\label{wdelta2}\\
w_{T',1bs}^{AV}
&=&
2\frac{f^2}{m_{\pi}^{2}}
\frac{G_{A}F_1^V}{q} 
\int \frac{d^3k}{(2\pi)^3}
\left(
\frac{4\nq\cdot\nk_1}{\nk_1^2+m_{\pi}^2}+
\frac{4\nq\cdot\nk_2}{\nk_2^2+m_{\pi}^2}
\right),
\\
w^{AV}_{T',1b\pi}
&=&
-4\frac{f^2}{m_\pi^2}F_1^V\frac{G_A}{q}
\int \frac{d^3k}{(2\pi)^3}
\frac{4(\nq\times\nk_2)^2}{(\nk_1^2+m_{\pi}^2)(\nk_2^2+m_{\pi}^2)},
\label{mpi}
\\
w_{T',1b\Delta}^{AV}
&=&
-   \sqrt{ \frac32 } \frac49  \frac{ff^*}{m_\pi^2}
\frac{G_{A}   C_3^V}{m_N q}
\frac{1}{m_\Delta-m_N}
\int \frac{d^3k}{(2\pi)^3}
2\left(
\frac{3q^2k_1^2-(\nq\cdot\nk_1)^2}{\nk_1^2+m_{\pi}^2}+
\frac{3q^2k_2^2-(\nq\cdot\nk_2)^2}{\nk_2^2+m_{\pi}^2}
\right),
\label{wdelta3}\\
w_{T',m\Delta}^{VA}
&=&
-\sqrt{ \frac32 } \frac49  \frac{ff^*}{m_\pi^2}
\frac{G_M^VC_5^A}{m_N q}
\frac{1}{m_\Delta-m_N}
\int \frac{d^3k}{(2\pi)^3}
2\left(
\frac{3q^2k_1^2-(\nq\cdot\nk_1)^2}{\nk_1^2+m_{\pi}^2}+
\frac{3q^2k_2^2-(\nq\cdot\nk_2)^2}{\nk_2^2+m_{\pi}^2}
\right).
\label{wdelta4}
\end{eqnarray}
\end{widetext}

%------------------
\section{Results}
%---------------------

In this section, we present numerical results for the effect of
two-body MEC on the charged-current (CC) neutrino response functions
in the 1p1h channel.  We consider three nuclear models: the
RFG, the RMF in
nuclear matter, and the SuSAM* approach.
 These results allow us to assess the model
dependence of the MEC contributions. Finally, we illustrate the impact
of these effects on neutrino cross sections with selected examples.
To this end, we will first examine in detail the interference terms
between the one-body and two-body currents, which provide the dominant
MEC contribution, and later compare them with the pure one-body
responses.

\begin{figure} %figure 3
  \centering
\includegraphics[width=8.5cm,bb=65 520 490 810]{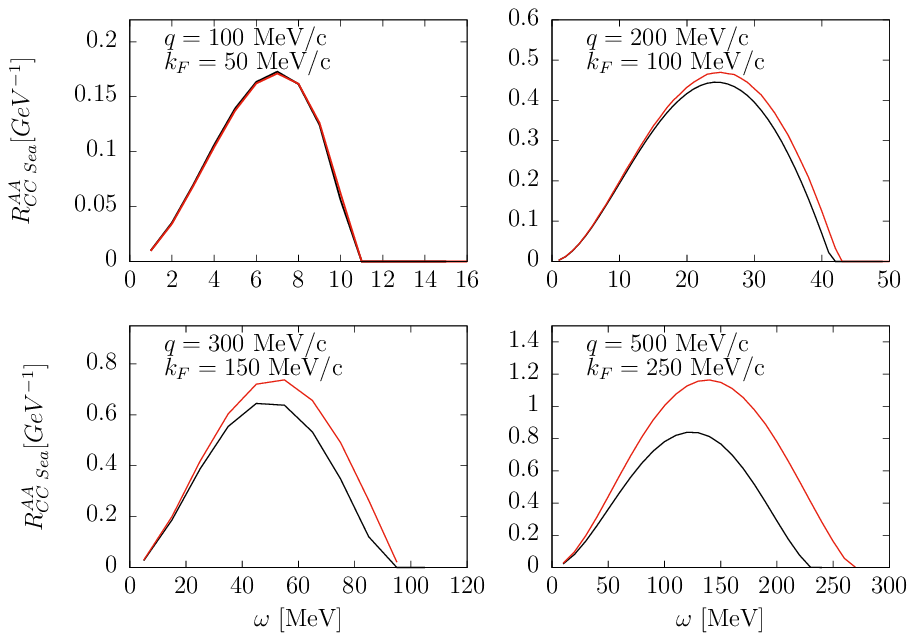}
\caption{Interference response $R_{CC}^{AA}$ between the one-body
  axial current and the seagull current, for increasing values of the
  momentum transfer $q$, with $k_F = q/2$. In each panel, the
  non-relativistic results (red lines) are compared with the
  relativistic ones (black lines).}
\label{relneu7}
\end{figure}

\begin{figure} % figure 4
  \centering
\includegraphics[width=8.5cm,bb=65 520 490 810]{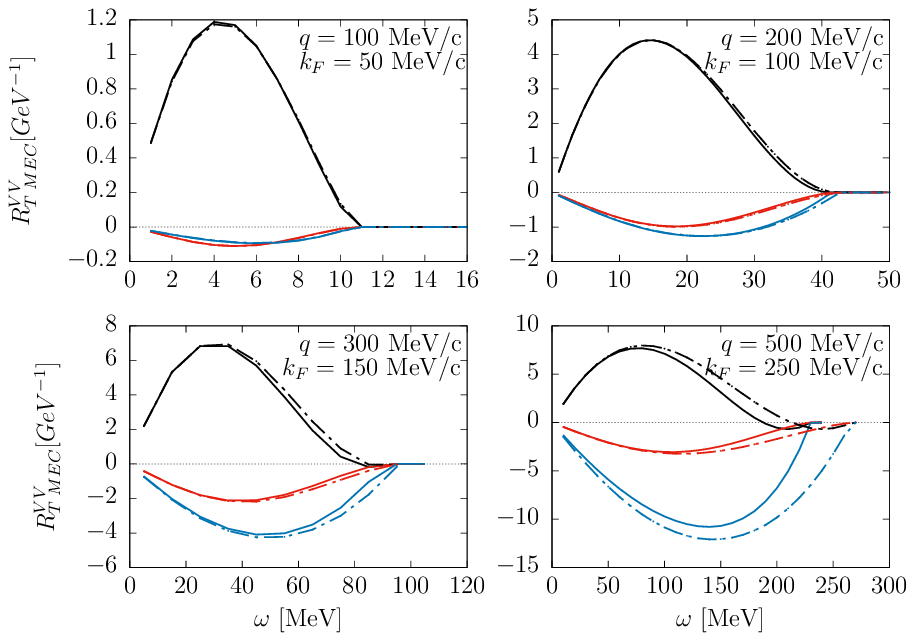}
  \caption{ Non-relativistic (dot-dashed lines) versus relativistic
    (solid lines) transverse interference response $R_T^{VV}$
    between the 1b current and the seagull (black),
    pion-in-flight (red), and $\Delta$-excitation (blue) two-body
    currents, for increasing values of $q$ and with $k_F = q/2$.  }
\label{relneu8}
\end{figure}

\begin{figure} % figure 5
  \centering
\includegraphics[width=8.5cm,bb=65 520 490 810]{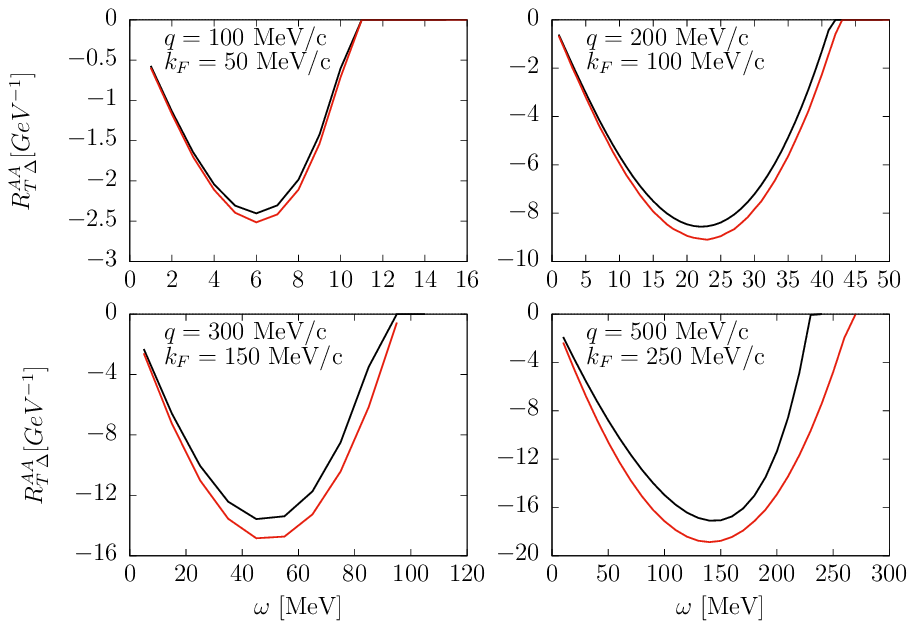}
\caption{The same as Fig. \ref{relneu7} for the
  interference response $R_T^{AA}$ between the axial 1b and $\Delta$
  currents}
  \label{relneu9}
\end{figure}

\begin{figure} % figure 6
  \centering
\includegraphics[width=8.5cm,bb=60 520 469 810]{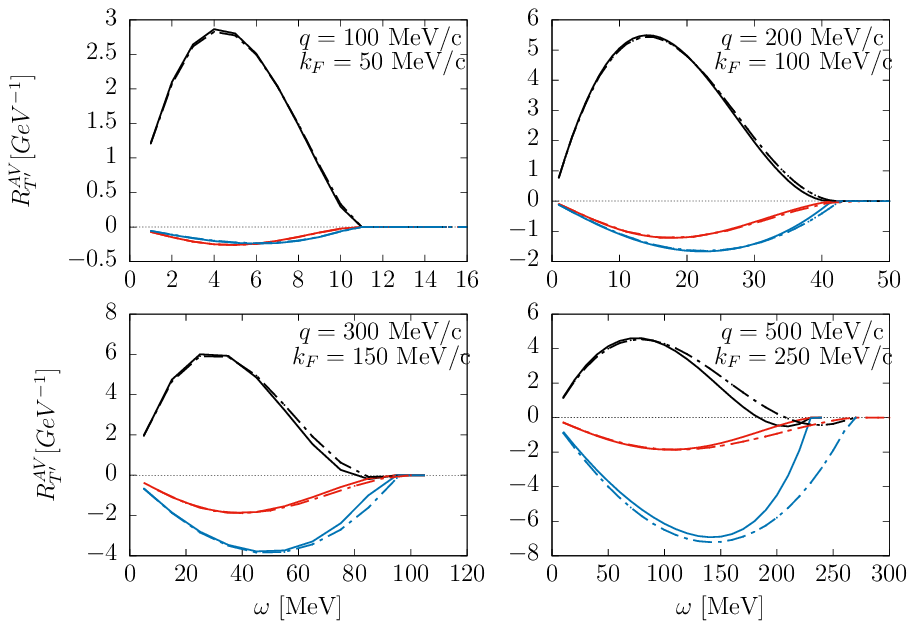}
\caption{The same as fig. \ref{relneu8} for the interference response
  $R_{T'1b2b}^{AV}$ between the axial 1b current and the
  vector seagull (black), pionic (red) and Delta (blue) currents.}
\label{relneu10}
\end{figure}

\begin{figure} % figure 7
  \centering
\includegraphics[width=8.5cm,bb=50 520 469 800]{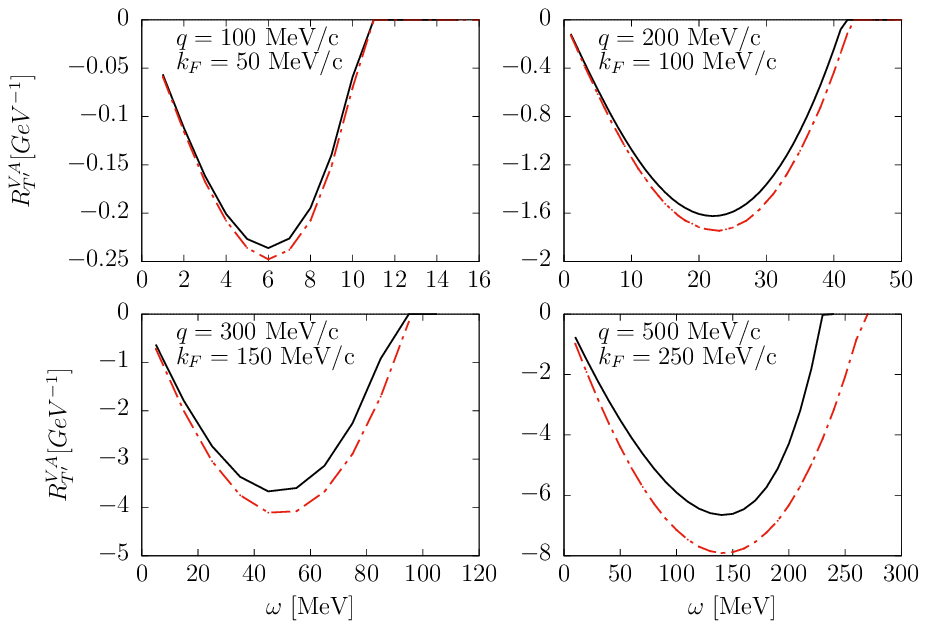}
\caption{The same as fig. \ref{relneu7}
  for the $R_{T' 1b \Delta}^{VA}$ interference response.}
\label{relneu11}
\end{figure}

\begin{figure*}  % figure 8
  \centering
\includegraphics[width=12cm,bb=20 331 550 790]{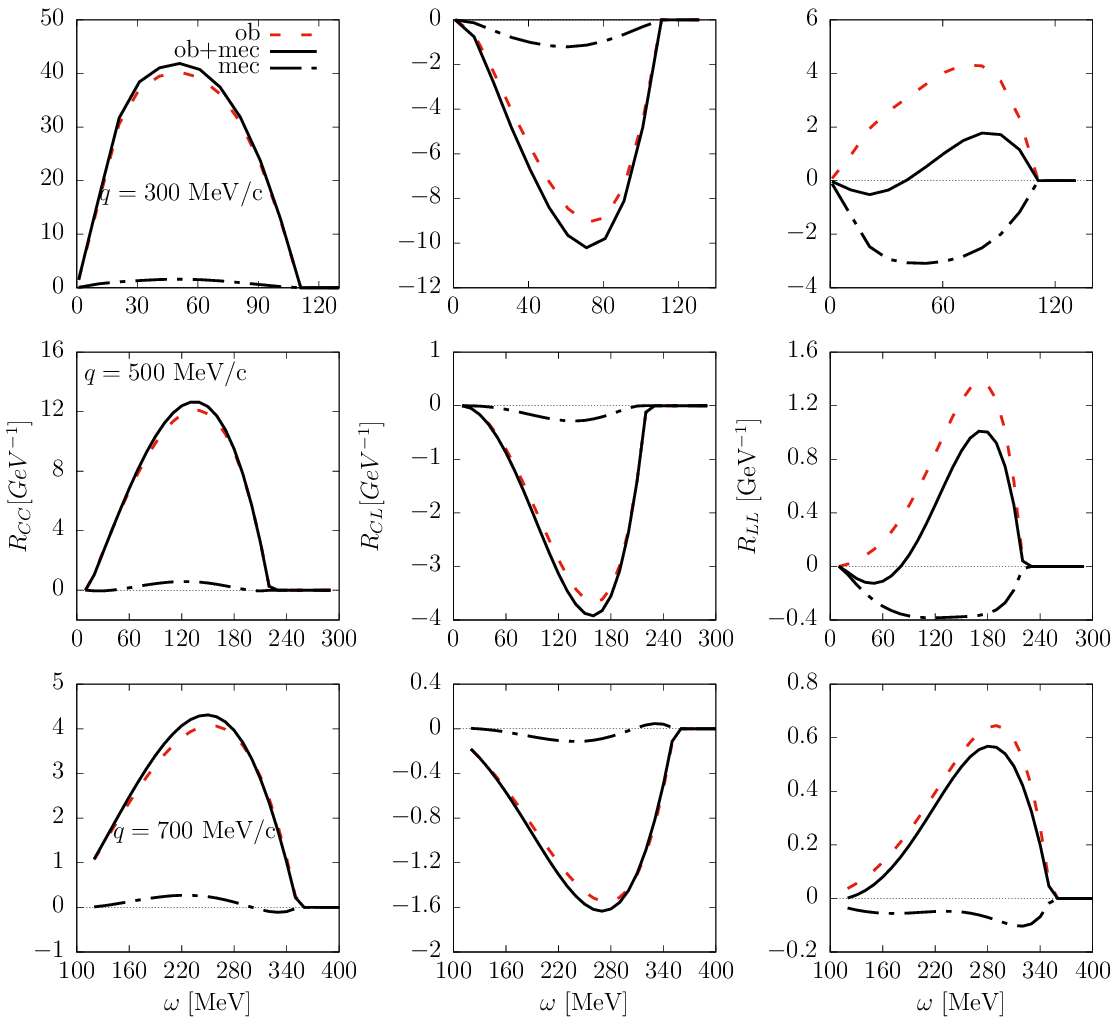}
  \caption{Response functions \(R_{CC}\), \(R_{CL}\),
    and \(R_{LL}\) computed in the RFG model for different values of $q$.
We compare one-body responses
with the MEC contribution and the total results.}
    \label{neufig7}
\end{figure*}

\subsection{Non-relativistic Fermi gas results}

We begin by analyzing the dominant response functions to leading order
in the non-relativistic Fermi gas (NRFG), and comparing them with the
fully relativistic results obtained within the RFG model. As
previously discussed, this comparison serves as a consistency test of
the calculation, since the NRFG and RFG models are implemented in
different ways. In the non-relativistic case,
the integrals over the momentum of the intermediate nucleon
are performed analytically following the method of Refs.~\cite{Ama94a,Ama94b},
except for the pion-in-flight current, where they reduce to
one-dimensional integrals. In addition, the spin traces have been
computed explicitly in the NR case. On the other hand, the RFG
responses are evaluated fully numerically. As expected, the RFG
results approach those of the NRFG in the low-$q$, low-$\omega$
region, as we explicitly demonstrate below.

In Fig.~\ref{relneu7} we show the interference response
$R_{CC,1b\,s}^{AA}$, which corresponds to the interference between the
axial one-body current and the axial seagull two-body current in the
CC channel. This is the dominant MEC contribution to the CC responses
at low momentum transfer. To examine how the relativistic response
approaches the non-relativistic one as $q$ becomes small, we present
results for several values of the momentum transfer: $q=100$, 200,
300, and 500 MeV$/c$. In each panel we set the Fermi momentum to
$k_F= q/2$, so that the momenta of the initial nucleons are also small
when $q$ is small. This choice also minimizes the effects of Pauli
blocking in the comparison. As expected, we observe that the
relativistic and non-relativistic responses are nearly identical for
$q = 100$ MeV$/c$, and begin to differ progressively as $q$ increases.

In Fig.~\ref{relneu8} we show the interference contributions to the
$R_T^{VV}$ response, separating the effects of the seagull,
pion-in-flight, and $\Delta$ currents. These vector-vector
responses are exactly twice as large as the corresponding electromagnetic
responses, as proven in Appendix~B. We observe that the seagull
contribution is positive, while the pion-in-flight and $\Delta$
contributions are negative, leading to a partial cancellation. The
$\Delta$ term increases with $q$ more rapidly than the seagull term,
and becomes the dominant contribution at $q = 500$~MeV$/c$. This
behavior is consistent with the electromagnetic response results
reported in Ref.~\cite{Cas25}.

In Fig.~\ref{relneu9} we present the axial transverse interference
response $R_T^{AA}$. At $q = 100$~MeV$/c$, the relativistic and
non-relativistic results are nearly identical, serving as a triple
consistency check: first, of the non-relativistic reduction of the
axial $\Delta$ current performed in Appendix C; second, of the
numerical implementation of the RFG and NRFG frameworks; and third, of
the analytic and numerical procedures used in each calculation. The response
is negative and its absolute value increases with $q$, becoming
slightly larger than the corresponding vector $\Delta$ contribution.

In Fig.~\ref{relneu10} we show the interference responses
$R_{T',1b2b}^{AV}$ between the axial one-body current and the
vector MEC, separating the contributions of the vector seagull,
pionic, and $\Delta$ currents. These responses exhibit a behavior very
similar to the corresponding $R_{T,1b2b}^{VV}$ ones. In fact,
by inspecting the expressions for the non-relativistic single-nucleon
responses, one finds that $w_{T,m2b}^{VV}$ (interference with
magnetization current) and $w_{T',1b2b}^{AV}$ involve exactly
the same integrals over the intermediate nucleon momentum $\nk$. The
difference lies in the coupling factors between the axial current and
the magnetization current. Since the convection current has little
impact on the transverse responses, the resulting behavior is nearly
identical in both cases.

Finally, in Fig.~~\ref{relneu11} we show the
$R_{T',1b\Delta}^{VA}$ interference response. Once again, we observe a close
similarity between this response and both
$R_{T',1b\Delta}^{AV}$ and $R_{T,1b\Delta}^{VV}$,
for the same reasons discussed previously. This similarity originates
from the fact that the structure of the $\Delta$ current and the
magnetization current is nearly the same in the axial and vector
sectors. The main difference lies in a vector product with $\vec{q}$,
while the spin operator is identical in both cases, except for
different coupling constants associated with the axial and vector
currents. This can also be seen by inspecting the corresponding
single-nucleon tensors, which involve the same integrals in all three
cases.

In all previous results, the sign of the one-body-$\Delta$
interference responses is always found to ne negative. This result can
be clearly understood by inspecting the expressions of the
single-nucleon response functions, Eqs. (\ref{wdelta1}),
(\ref{wdelta2}), (\ref{wdelta3}) and (\ref{wdelta4}). All of them
carry an overall minus sign and are proportional to the same integral,
which is positive since the integrand contains terms of the form \(
3q^2k_1^2 - (\mathbf{q} \cdot \mathbf{k}_1)^2 > 0 \). This result for
the sign of the \(1b\)--\(\Delta\) interference was already pointed
out in Ref.~\cite{Cas25} for the electromagnetic transverse
response. A similar argument applies to the \( R^{VV}_{T,m\pi} \) and
\( R^{AV}_{T',1b\pi} \) responses, which are also negative. Although
these sign arguments are strictly valid in the non-relativistic limit,
numerical calculations show that they remain valid in the relativistic
case, except at very large \( q \), where sign changes can occur
depending on the value of \( \omega \).

\begin{figure} % figure 9
  \centering
\includegraphics[width=8cm,bb=80 330 447 800]{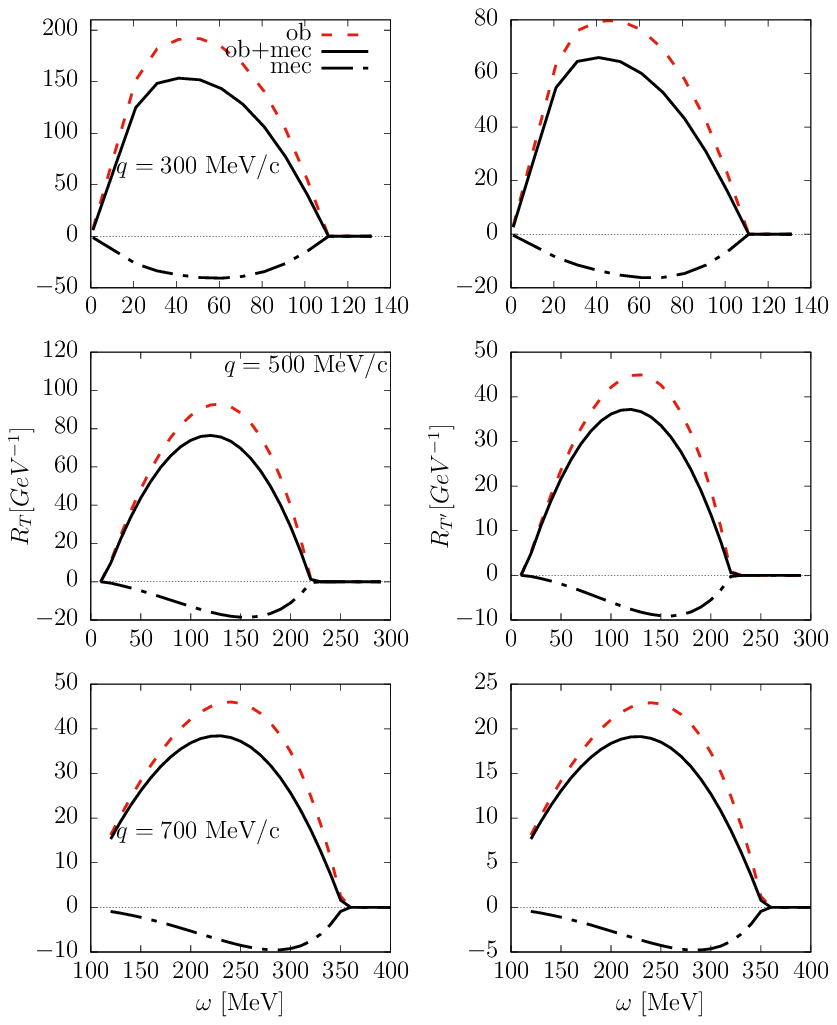}
 \caption{The same as in Fig. \ref{neufig7} for the transverse responses
   $R_T$ and $R_{T'}$}
 \label{neufig11}
\end{figure}

\begin{figure}  % figure 10
  \centering
 \includegraphics[width=8cm,bb=85 320 450 810]{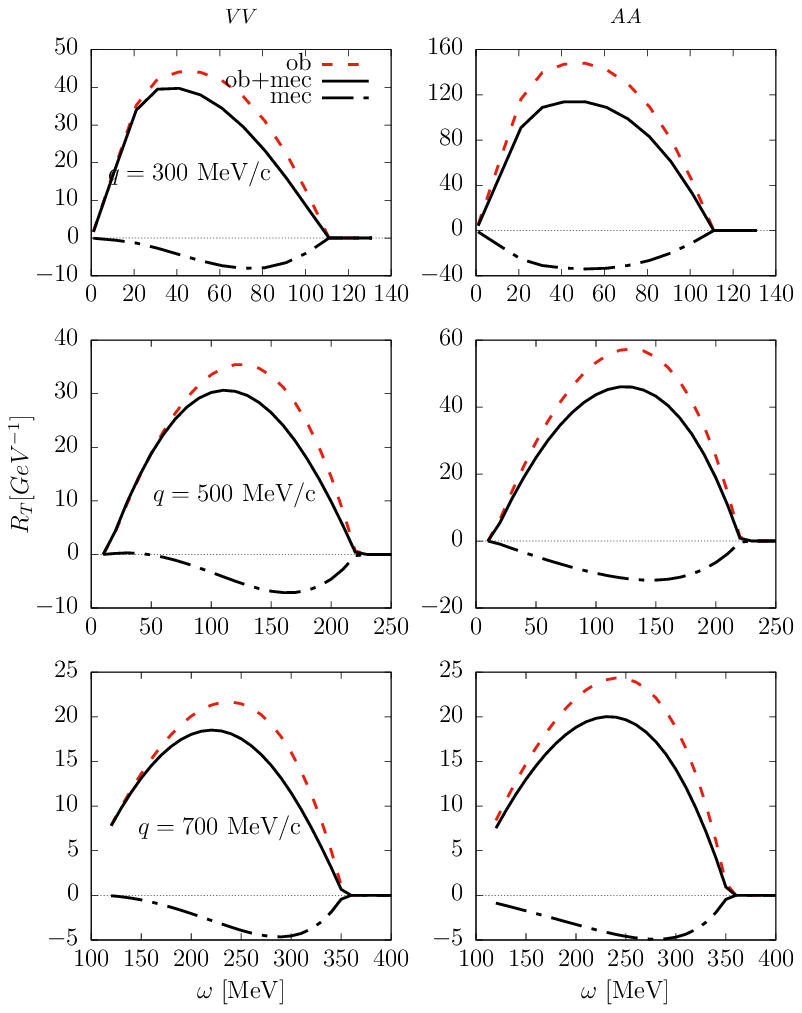}
 \caption{ The same as fig. \ref{neufig11} for the $R_T^{VV}$ and $R_T^{AA}$
   response functions.
   }
 \label{neufig9}
\end{figure}

\begin{figure}   % figure 11 
  \centering
\includegraphics[width=8cm,bb=80 340 445 800]{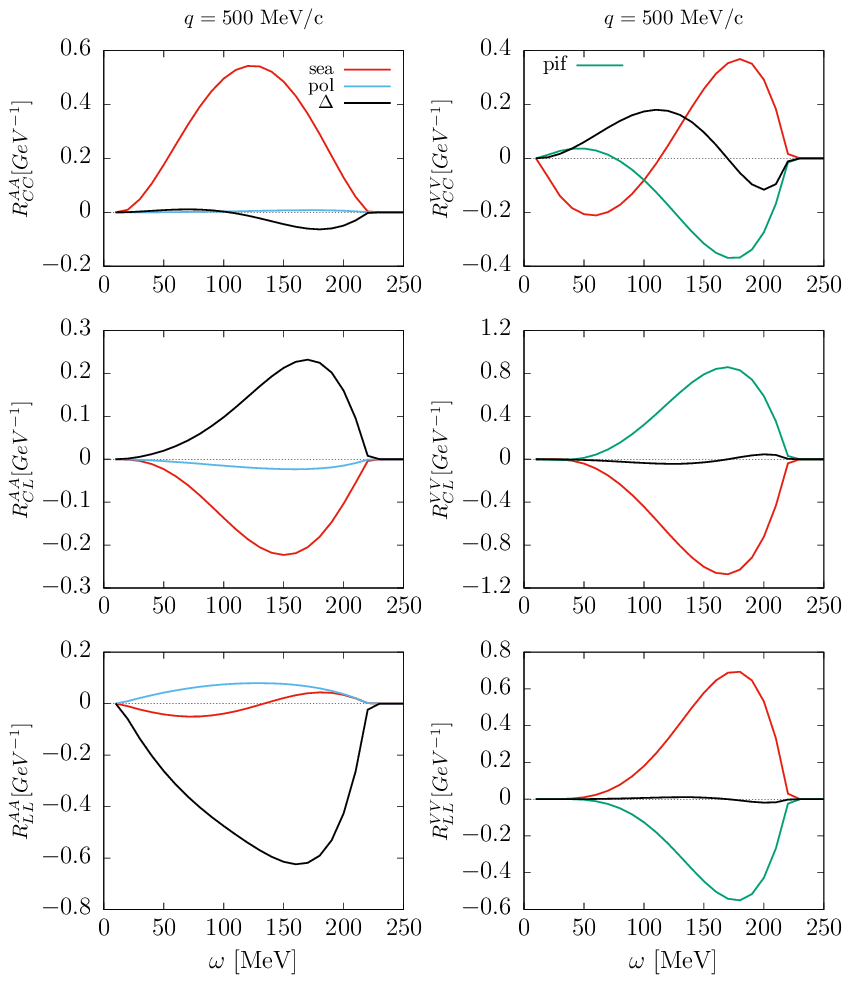}
 \caption{Interference longitudinal responses between the 1b current
   and the different MEC seagull (sea), pionic (pif), pion pole (pol)
   and $\Delta$, separated in vector and axial contributions, for
   $q=500$ MeV/c.  }
 \label{neufig13}
\end{figure}

\begin{figure}   % figure 12
  \centering
 \includegraphics
[width=8cm,bb=85 480 450 810]{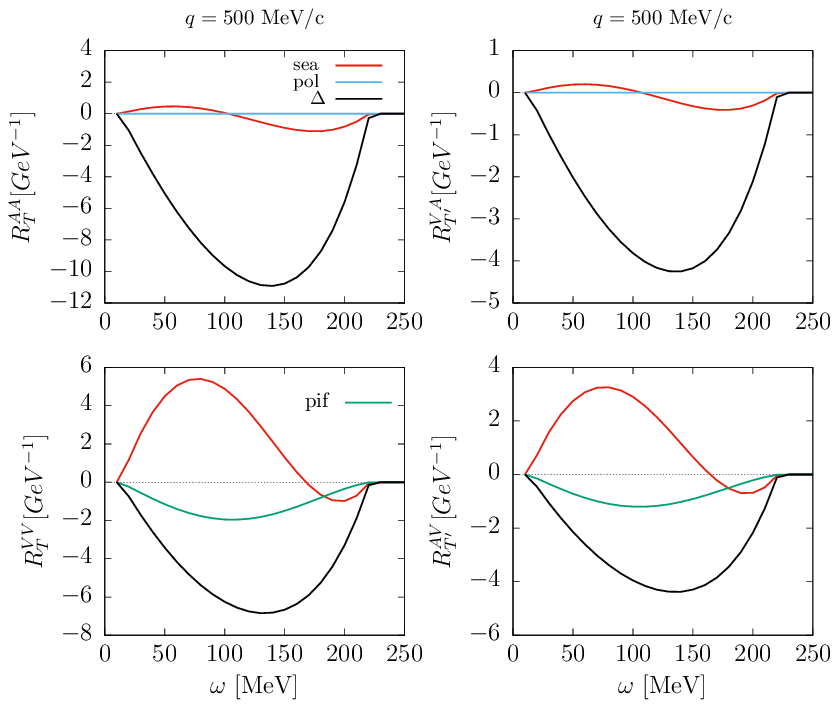}
 \caption{The same as Fig. \ref{neufig13} for the transverse responses }
 \label{neufig16}
\end{figure}

\subsection{Results in the RFG model}

We now introduce the results obtained within the RFG model. In this
section, we apply the formalism to the case of $^{12}$C, using a Fermi
momentum \( k_F = 225 \) MeV/c.

In Fig.~\ref{neufig7} we show the longitudinal responses \( R_{CC} \),
\( R_{CL} \), and \( R_{LL} \), while in Fig.~\ref{neufig11} we
display the transverse responses \( R_T \) and \( R_{T'} \). For each
response, we present three curves: the pure one-body (1b)
contribution, the full result including MEC (1b+mec), and the
interference term between one-body and two-body currents (mec). It is
worth noting that the full results comprise both the interference term
and the (small) standalone contribution from MEC.

It is observed that the effect of MECs in the \( R_{CC} \) response is
a very small increase. In the \( R_{CL} \) response, MECs introduce a
visible modification, and in the \( R_{LL} \) response they are of the
same order of magnitude as the one-body part for small values of
$q$. Nevertheless, this will have little or no impact on neutrino
cross sections, since \( R_{LL} \) is very small and longitudinal
responses in general contribute less significantly than the transverse
ones to the total cross section.

Figure~\ref{neufig11} shows the transverse responses \( R_T \) and \(
R_{T'} \) in the relativistic Fermi gas model. The interference
is negative and leads to a significant reduction of both
responses. This reduction was also found in the electromagnetic
transverse response \cite{Cas25}.
This behavior is due to the interference
with the \(\Delta\) current that 
 gives a dominant contribution in the
 relevant kinematics.

 The MEC-induced suppression of the transverse neutrino responses is
 of similar magnitude in the vector and axial sectors.  To illustrate
 this point, in Fig.~\ref{neufig9} we show separately the
 vector-vector (\( VV \)) and axial-axial (\( AA \)) contributions to
 the transverse response \( R_T \). It can be seen that the reduction
 induced by MEC, relative to the one-body current, is similar in both
 channels. In both cases, the dominant effect is the interference
 between the one-body current and the \(\Delta\) current, which leads
 to a negative contribution.

A more detailed scrutiny of the relevance of each MEC contribution to
the different response functions is provided in Figs. \ref{neufig13}
and \ref{neufig16}, where all the interference terms are shown
separately for the seagull, pionic, pion-pole, and \(\Delta\)
currents. The effect of MECs on the longitudinal responses is diverse,
Fig. \ref{neufig13}, although these interferences have a limited
impact on the total responses and an even smaller one on the neutrino
cross section. For example, the pionic and seagull contributions tend
to cancel each other in the VV-type \(R_{CC}\), \(R_{CL}\), and
\(R_{LL}\) responses. In the axial \(R_{CC}\), the seagull clearly
dominates, as anticipated in the non-relativistic development, while
in the axial \(R_{CL}\) and \(R_{LL}\) responses a non-negligible
contribution from the \(\Delta\) current is observed, since the axial
\(\Delta\) is no longer purely transverse. The pion pole is
negligible.  Nevertheless, we reiterate that the overall impact of
these longitudinal interferences on observables is minimal. The MEC
effect that plays a significant role appears in the transverse
responses of Fig. \ref{neufig16}, where the \(\Delta\) current is
clearly dominant and produces a sizable reduction of the response,
partially compensated by the seagull contribution.

An important outcome of this analysis is the identification of a
sizeable negative interference between the one-body and
$\Delta$-current contributions in the axial channel, particularly in
the transverse and axial-transverse responses. This effect, which to
our knowledge has not been previously highlighted in the literature on
quasielastic neutrino scattering, is a novel result of this work. It
is especially relevant for neutrino interactions, where the axial
current plays a central role. Since current neutrino event generators
and models often neglect such interference terms and treat MEC
contributions only as a 2p2h additive term, this finding suggests that
existing models may require revision in order to properly account for
interference effects, especially those involving the $\Delta$ current
in axial channels.

\begin{figure} % figure 13
  \centering
\includegraphics[width=8.5cm,bb=60 280 490 810]{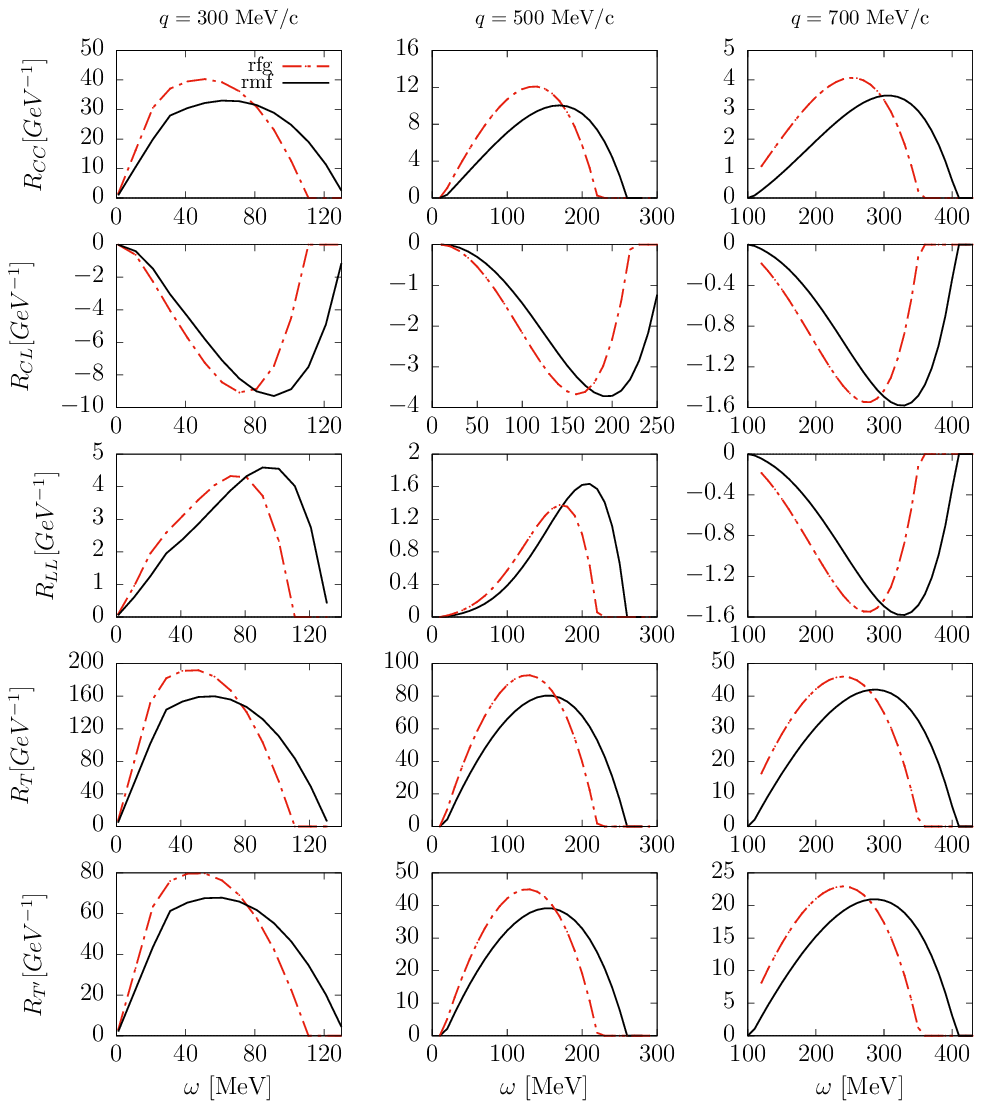}
\caption{Comparison of the five total (1b+MEC) response functions $R_{CC}$,
  $R_{CL}$, $R_{LL}$, $R_T$, and $R_{T'}$ computed within the
  RFG and the RMF
  models for momentum transfers $q = 300$, 500, and 700 MeV$/c$.
}
 \label{neufig8}
\end{figure}

\begin{figure}  % figure 14
  \centering
 \includegraphics[width=8cm,bb=75 340 450 810]{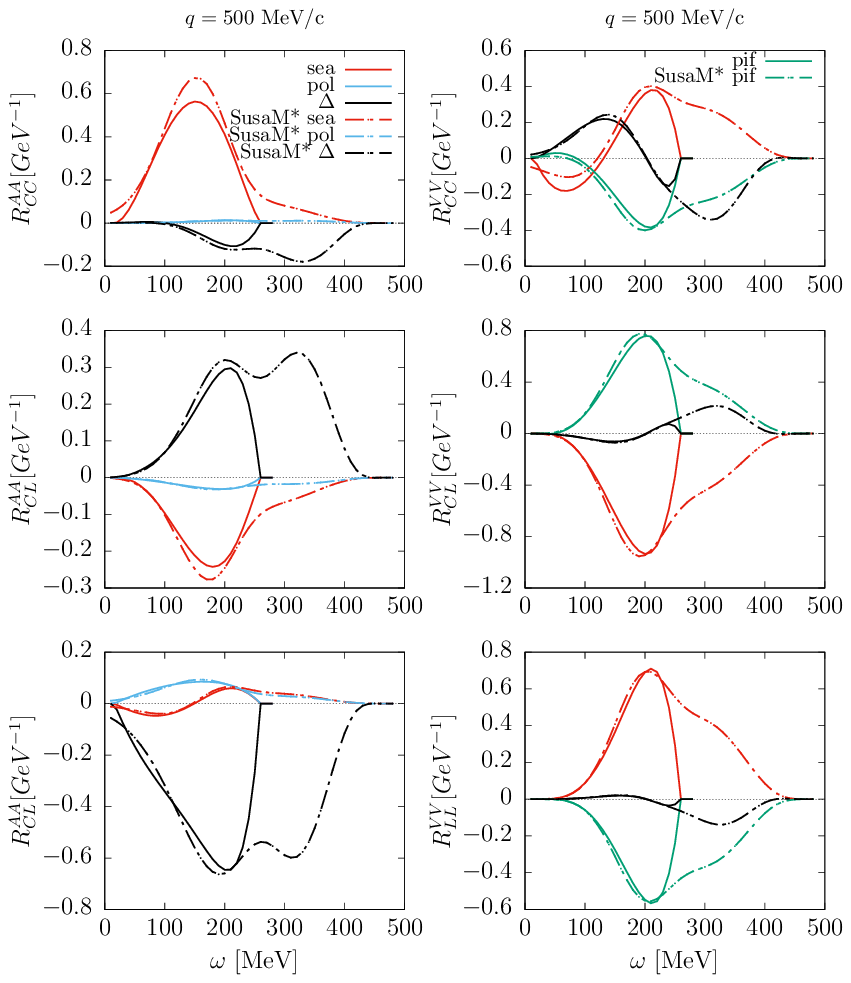}
 \caption{Longitudinal 1b-MEC interference responses in the SuSAM* approach
   approach compared to the RMF model for $q=500$ MeV/c.}
 \label{neufig2}
\end{figure}

\begin{figure}   % figure 15
  \centering
\includegraphics[width=8cm,bb=75 480 445 810]{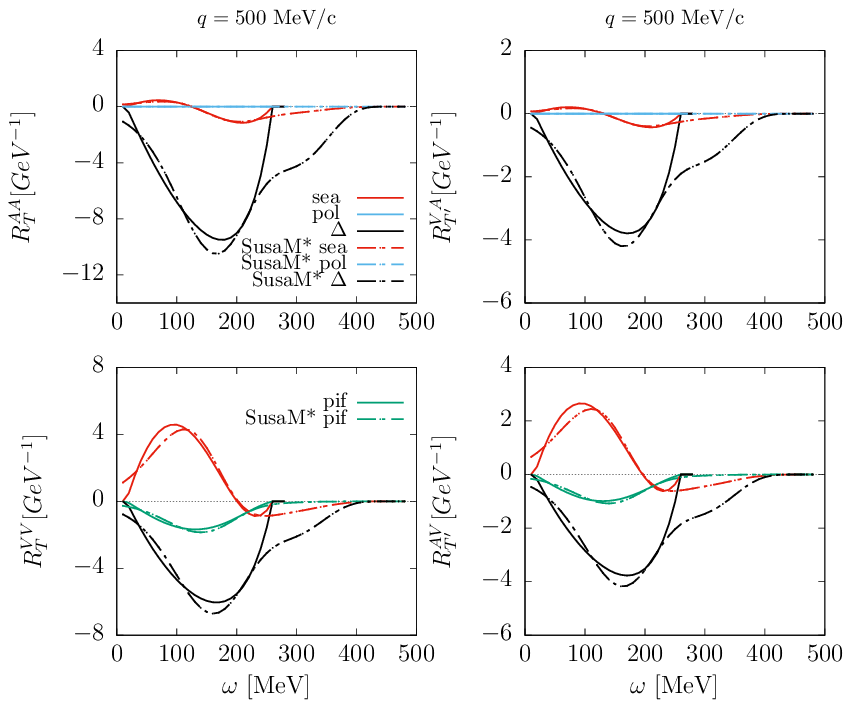}
 \caption{The same as Fig.(\ref{neufig2}) for the transverse interference
   responses. }
 \label{relneu6}
\end{figure}

\begin{figure*} % figure 16
  \centering
\includegraphics[width=10cm,bb=50 295 490 800]{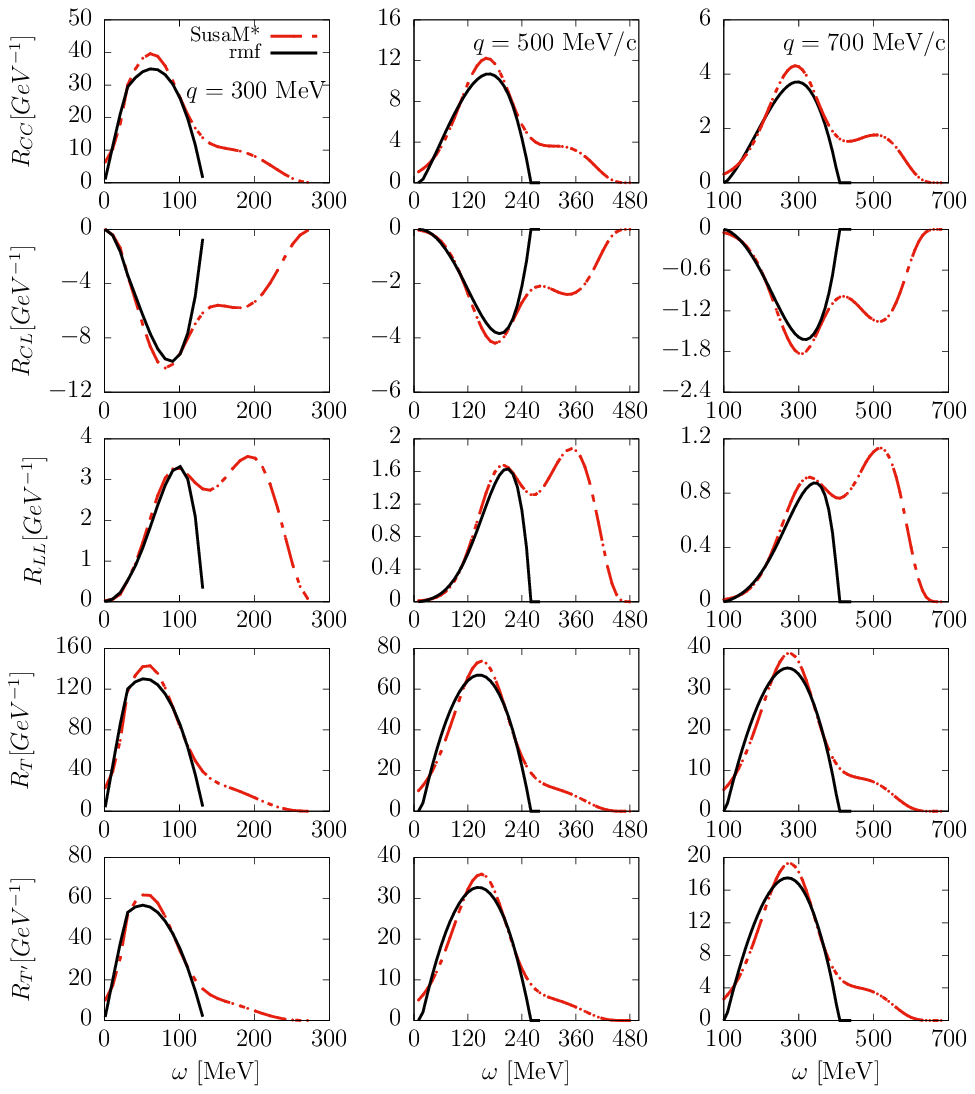} 
 \caption{Total responses in the RMF and SuSAM* models for momentum
   transfers $q=300,500,700$ MeV/c.}
 \label{relneu12}
\end{figure*}

\subsection{Relativistic mean field and superscaling}

In this subsection we present results obtained with two additional
nuclear models beyond the RFG:
the RMF model of nuclear matter and the
superscaling analysis with relativistic effective mass.

\subsubsection{RMF}

In this model, nucleons interact via a relativistic constant mean
field composed of scalar and vector potentials, as described in the
Walecka model \cite{Kim94}. As a result, nucleons acquire an effective
mass $m_N^*$ due to their coupling to the scalar field, along with a
constant vector potential energy $E_v$. The single-particle energy of
nucleons in the RMF is thus given by
\begin{equation}
E_{\text{RMF}} = \sqrt{p^2 + (m_N^* )^2} + E_v.
\end{equation}
The formalism for computing the nuclear response functions is modified
accordingly: the on-shell energy $E = \sqrt{p^2 + m_N^{*2}}$ is used
in the kinematics, and the free Dirac spinors $u(p)$ are replaced by
spinors constructed with the effective mass $m_N^*$.
Therefore all the formulas from the previous section can be applied
 but replacing $m_N$ by  $m_N^*$ everywhere except in the current
operators, which remain unmodified in this model.
Additionally, the vector potential energy
$E_v$ must be included in the $\Delta$ propagators to
compute the four momenta $p+q$ and $p'-q$.
With these prescriptions, the electromagnetic response functions
including MEC in the 1p1h channel were
evaluated in Ref.~\cite{Cas23}. We now extend this framework to the
weak sector, focusing on the role of MEC and interference effects in
neutrino-nucleus reactions.

In Fig.~\ref{neufig8}, we compare the total five response functions,
including 1b plus MEC contributions,
calculated in the RFG and RMF models for three values of the momentum
transfer. The main effect of the relativistic mean field is to produce
a shift of the strength towards higher values of $\omega$
due to the use of an effective mass for the
nucleons. This shift effectively incorporates, through the mean field,
part of the dynamical effects related to the binding and interaction
energy of the nucleon in the final state.

\subsubsection{SuSAM*}

The SuSAM* model  combines the superscaling
approach with elements of the RMF model through the introduction of an
effective mass. This approach is based on the assumption that the
response functions can be factorized as the product of an averaged
single-nucleon response and a phenomenological superscaling function,
extracted from electron scattering data~\cite{Cas23}. The SuSAM* model
provides an alternative to the SuSA approach, using a single universal
scaling function for all channels, and it allows for the inclusion of
MEC effects both in the generalized
single-nucleon responses and in the scaling function itself. The model
was originally developed to describe electromagnetic responses, and it
is now extended to the weak sector including MEC contributions.

The starting point is to  rewrite the RMF response functions,
Eq.~(\ref{respuesta}) in a factorized form by defining:
\begin{itemize}
\item  an averaged
single-nucleon response $\overline{w_K}$
\begin{equation}\label{averaged}
  \overline{w_K}(q,\omega)=
  \frac
      { \int_{\epsilon_0}^\infty d\epsilon\; n(\epsilon) w_K(\epsilon,q,\omega)}
{ \int_{\epsilon_0}^\infty d\epsilon\; n(\epsilon)},
\end{equation}
\item 
a scaling function
\begin{equation}
  f^*(\psi^*)=\frac34 \frac{1}{\epsilon_F-1}
   \int_{\epsilon_0}^\infty d\epsilon\; n(\epsilon),
\end{equation}
\item
  and a scaling variable
  \begin{equation}
    \psi^* = \sqrt{\frac {\epsilon_0-1}{\epsilon_F-1}}
      \mbox{sgn}(\lambda-\tau)
  \end{equation}
\end{itemize}
Using $V/(2\pi^3)=N/(\frac83\pi k_F^3)$, were $N$ is the number of neutrons,
the responses can be written in the factorized form
\begin{equation} \label{susam}
  R_K(q,\omega)= \frac{\epsilon_F-1}{m_N^* \eta_F^3 \kappa}N\overline{w_K}(q,\omega)f^*(\psi^*) 
\end{equation}
where $\eta_F=k_F/m_N^*$.
In the RMF (and also the RFG in the particular case $m_N^*=m_N$) 
the condition $\epsilon_0 < \epsilon_F$ restricts the kinematical
region in $\omega$ where the response is non-zero, corresponding to
the interval $-1 < \psi^* < 1$ in terms of the scaling variable.
In the SuSAM* model, this region is
extended beyond the RFG boundaries using a phenomenological scaling function,
$f^*_{\rm ph}(\psi^*)$,
extracted from inclusive electron scattering data.

\begin{figure*}
  \centering
\includegraphics[width=14cm,bb=15 310 530 785]{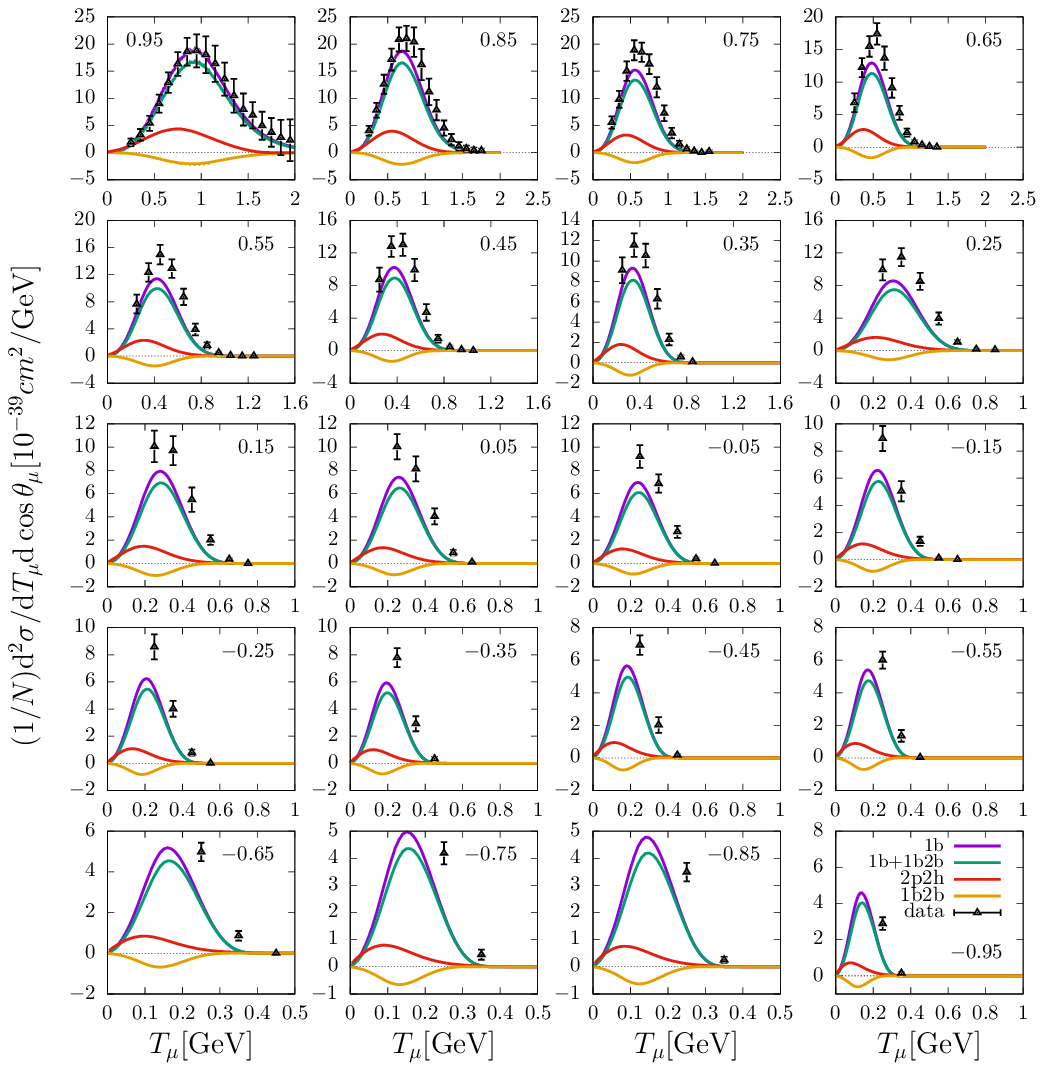}

\caption{Flux-integrated double-differential cross section per neutron
  for CC neutrino scattering on $^{12}$C in the RMF model. The
  experimental points are the inclusive CCQE measurements from
  MiniBooNE~\cite{Agu10}. Shown are the one-body (1b) results, the
  1b--2b interference (1b2b), their sum (1b+1b2b), and the 2p2h
  contribution from Ref.~\cite{Mar23} computed within the same RMF
  model.}

 \label{neutrinocs}
\end{figure*}

However, there is the problem that the averaged single-nucleon in Eq
(\ref{averaged}) is not defined for $\epsilon_0> \epsilon_F$ (or
$|\psi^*|>1$) because the denominator is zero.  To apply the model
beyond the RFG limits, one must extrapolate the averaged
single-nucleon responses outside the physically allowed region (as in
the SuSA approach), which is not feasible when MEC are included, unless an explicit analytical expression is
available. The solution adopted in Ref.~\cite{Cas23} was to slightly
modify the momentum distribution of the Fermi gas by introducing a
smooth smearing function of Fermi type, which allows all the values of
$\epsilon_0$
\begin{equation}
n(p) = \frac{1}{1 + \exp\left[(p - k_F)/a\right]},
\end{equation}
where $k_F$ is the Fermi momentum and $a \approx 25$ MeV is a diffuseness parameter
that controls the smoothness of the fall-off. This modification allows
one to compute the averaged single-nucleon responses $\overline{w_K}$
in a broader kinematic domain and to include the contribution of MEC
in a consistent way (see \cite{Cas23b,Cas23} for details).

In the SuSAM* approach, we use Eq.~(\ref{susam}) together with the
phenomenological scaling function extracted in Ref.~\cite{Cas23}. The
results for the interference $1b$--MEC response functions are shown in
Figs.~\ref{neufig2} and \ref{relneu6} for $q=500$ MeV$/c$. As we can
see, the interference responses extend well beyond the allowed region
of the RFG, enabling the estimation of MEC effects at large $\omega$
values. The phenomenological scaling function was parametrized as a
sum of two Gaussians, which explains why some of the responses display
two peaks. It is also apparent that in the transverse responses the
interference remains negative due to the $\Delta$ current.

In Fig.~\ref{relneu12} we compare the total response functions
computed in RMF and SuSAM*.  The longitudinal responses CL and LL are
small and contribute little to the neutrino cross section.  For the
dominant responses---the transverse ($T$, $T'$) and charge-charge
($CC$) ones---the effect of the SuSAM* model is to introduce a
high-energy tail in the responses. The relative size of the responses
can be clearly seen in this figure for $q=500$ MeV/c: the $T$ response
reaches a maximum of approximately $\approx 60$~GeV$^{-1}$ , $T'$ is
about half of that ($\approx 30$~GeV$^{-1}$), $CC$ peaks around
$\approx 10$~GeV$^{-1}$, $CL \approx -4$~GeV$^{-1}$, and
$LL \approx 1.5$~GeV$^{-1}$.
This shows that the $CC$ contribution is relatively
small, $CL$ even smaller, and $LL$ is almost negligible. Note,
however, that each response is weighted by a different kinematic
factor $v_K$ in the cross section, Eq. (\ref{cross}).
In particular, the $LL$ response is
so small that it becomes extremely sensitive to fine details of the
model, but such differences are likely to be unobservable in the total
cross section, which is largely dominated by the transverse responses
and, to a lesser extent, by the $CC$ component.

\subsection{Cross section}
%------------------------

\begin{figure*}
  \centering
\includegraphics[width=14cm,bb=15 510 530 790]{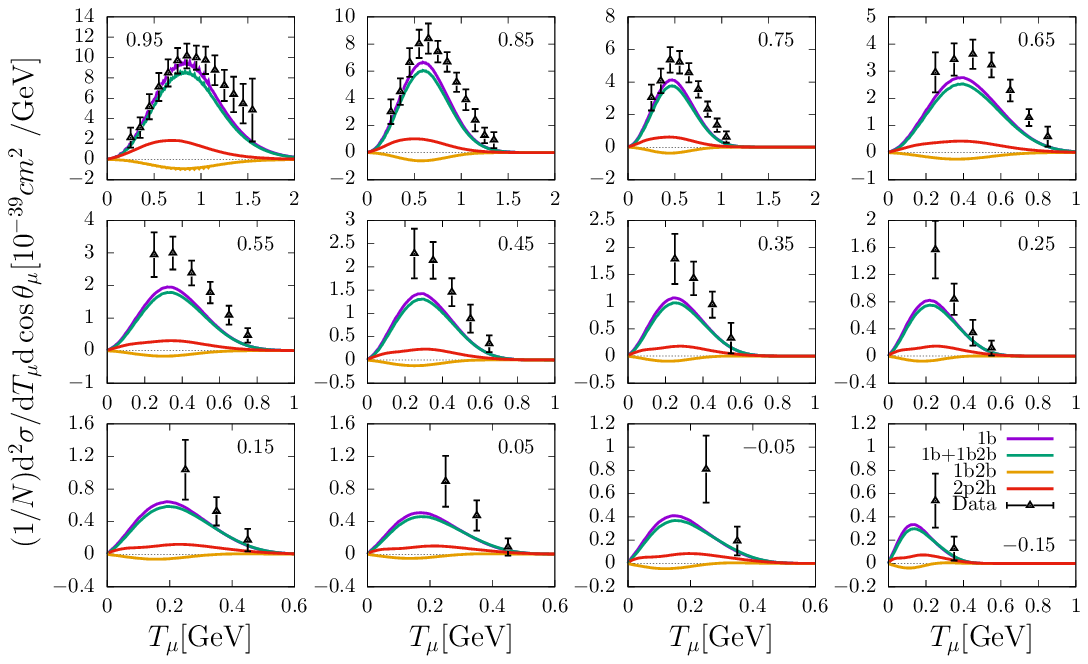}
\caption{ The same as Fig. \ref{neutrinocs} for the antineutrino cross
  section. Experimental data are the CCQE antineutrino meassurements
  from MiniBoone \cite{Agu13} }.
 \label{antineutrinocs}
\end{figure*}

In this final section we present results for the neutrino
$(\nu_\mu,\mu^-)$ and antineutrino $(\overline{\nu}_\mu,\mu^+)$
inclusive cross sections.  To compare theoretical predictions with
experimental data, the double differential cross section, expressed as
a function of the muon kinetic energy and the scattering angle, must
be integrated over the neutrino flux. The flux-averaged cross section
is defined as:
\begin{equation}
  \frac{d^2\sigma}{dT_\mu dcos\theta}= \frac{1}{\Phi_{tot}}\int dE_\nu \Phi(E_\nu) \frac{d^2\sigma}{dT_\mu dcos\theta}(E_\nu)
\end{equation}
where $\Phi(E_\nu)$ is the neutrino flux, $\frac{d^2\sigma}{dT_\mu dcos\theta}(E_\nu)$ is the cross section evaluated at a fixed neutrino energy $E_\nu$ and $\Phi_{tot}$  is the total integrated flux,
\begin{equation}
  \Phi_{tot} = \int dE_\nu \phi(E_\nu).
\end{equation}
The experimental data are typically provided in bins of
$\cos\theta$, where $\theta$ is the scattering angle of the outgoing
muon. For each bin, what is actually given is the cross section
averaged over the bin width, which implies an integration over
$\cos\theta$.
\begin{eqnarray}
  \left\langle \frac{d^2\sigma}{dT_\mu d\cos \theta} \right \rangle_{Bin}
& = &
  \nonumber\\
  && \kern -2.5 cm
\frac{1}{\Delta \cos \theta}
  \int_{\cos \theta_i}^{\cos \theta_f} \frac{d^2\sigma}{dT_\mu d\cos\theta} (\cos \theta) d \cos\theta.
\end{eqnarray}

In Figs.~\ref{neutrinocs} and \ref{antineutrinocs} we present results
for the double-differential charged-current neutrino and antineutrino
cross sections, respectively, corresponding to the kinematics and flux
of the MiniBooNE
experiment~\cite{Agu10,Agu13}.
The experimental data correspond to the fully inclusive CCQE
measurements, without any kinematical cuts, where the
CC1$\pi^+$ interaction with intranuclear pion absorption has
been subtracted.
Each
panel shows the cross section for a given $\cos\theta$ bin, with bin
width $\Delta\cos\theta = 0.1$, as a function of the kinetic energy of
the outgoing muon. A broad peak is observed, which arises from an
average over many cross sections corresponding to different values of
the incident neutrino energy $E_\nu$, weighted with the flux. This
averaging produces a much broader shape than what would be expected
from a quasielastic cross section at fixed $E_\nu$.

The theoretical calculations have been performed using the RMF model
with effective mass $M^* = 0.8$ and Fermi momentum $k_F = 225$ MeV/c,
corresponding to $^{12}$C. 

\begin{figure*}
  \centering 
\includegraphics[width=11.1cm,bb=15 300 530 690]{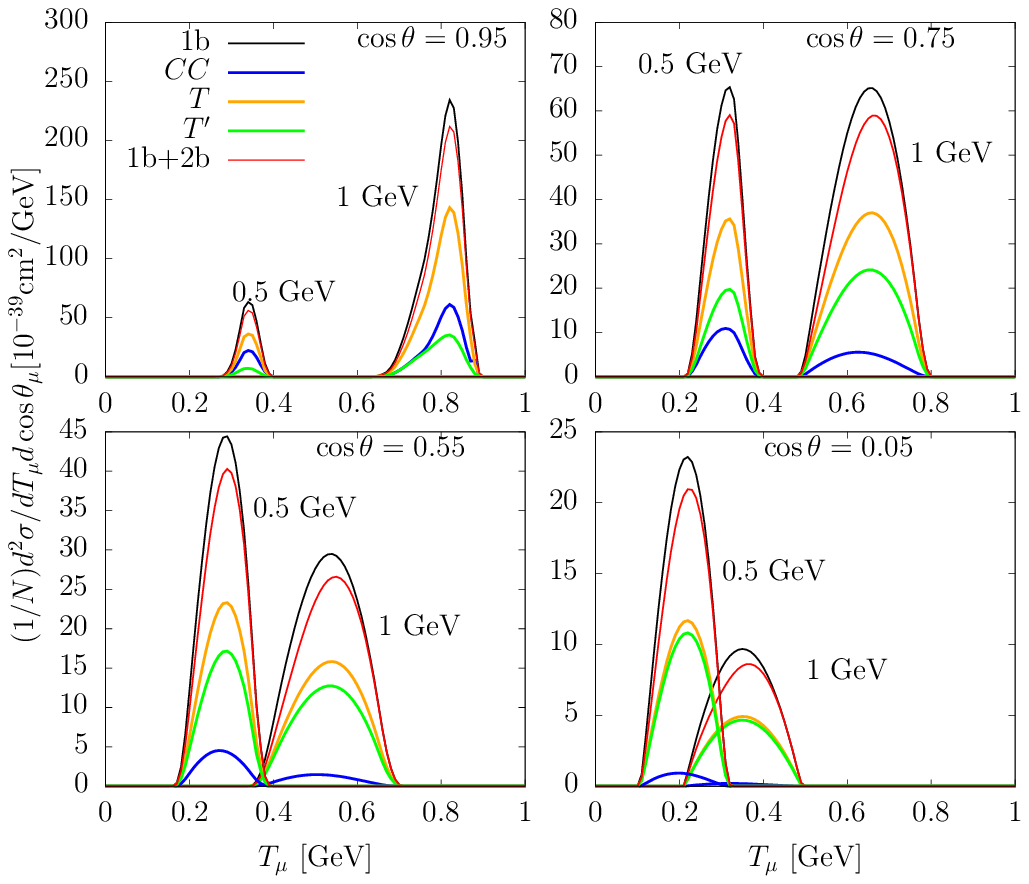}
\caption{  
Double-differential cross sections for CC neutrino scattering on
$^{12}$C in the RMF model at fixed neutrino energies $E_\nu=0.5$ and
1~GeV. Results are shown for different scattering angles corresponding
to $\cos\theta=0.95$, 0.75, 0.55, and 0.05. For the most forward bin,
$\cos\theta=0.95$, the cross section is averaged over
$\cos\theta\in[0.9,1]$, corresponding to angles between 0$^\circ$ and
25$^\circ$. In each panel we display the results with the one-body
current alone (1b) and including the interference contribution
(1b+1b2b). The 1b contributions of the $CC$, $T$, and $T'$ responses are
also shown separately.
}
 \label{figfinal}
\end{figure*}

In Figs.~\ref{neutrinocs} and \ref{antineutrinocs}
we show the one-body (1b) results together with the
sum of 1b and the 1b--2b interference. One observes that the
interference with MEC produces a reduction of the cross section of
about 10\%. For antineutrinos the effect is smaller due to the partial
cancellation between the $T$ and $T'$ responses. In the same figures
we also show separately the 1b--2b interference and the 2p2h
contributions for comparison. The 2p2h responses, computed in
Ref.~\cite{Mar23}, are positive, somewhat larger than the
interference, and display their maximum at lower energy transfer. As a
consequence, the two contributions partially cancel each other,
although the 2p2h still dominates. Therefore, one expects that when
both are summed the net effect would be an enhancement smaller than
that produced by the 2p2h contribution alone. We have not included
this sum in the figure, since the focus of the present work is on the
interference, but it is clear that both effects, 2p2h and 1b--2b
interference, are of comparable size.

Other models in the literature often neglect the interference between
one-body and two-body currents, and their predictions vary
significantly among them, contributing to the theoretical systematic
uncertainties in neutrino oscillation experiments. Our results
indicate that the 1b–2b interference constitutes an additional source
of uncertainty that should also be taken into account.

Overall, the agreement with experimental data is good at very forward
angle, while the theoretical calculation tends to underestimate the
data as $\theta$ increases.  This is expected given that the 2p2h
contributions have not been included yet.

In order to gain deeper insight into the flux–folded results of
Figs.~17 and 18, in Fig.~19 we show the impact of the interference
term on double-differential cross sections at fixed neutrino energies
and different scattering angles. Specifically, results are presented
for $\cos\theta=0.95,\,0.75,\,0.55$ and $0.05$. Because the cross
section varies rapidly at forward angles, the case $\cos\theta=0.95$
is averaged over the bin $\cos\theta\in[0.9,1]$, corresponding to
scattering angles between $0^\circ$ and $25^\circ$. For larger angles
the cross section varies smoothly within the bin, and it is sufficient
to show the result at the midpoint, corresponding to $41^\circ$,
$57^\circ$ and $87^\circ$, respectively. Each panel displays results
for incident neutrino energies of $0.5$ and $1$~GeV, which
approximately span the region where the flux is most intense. For each
kinematics we show the cross section obtained with the 1b current and
after including the interference (1b+2b). The results indicate that
the interference effect depends only weakly on kinematics, consistent
with the behavior observed in Figs.~17 and 18. One might expect that
the transverse response becomes more relevant at larger angles,
thereby increasing the relative effect of MEC. This is also illustrated in
Fig.~19, where the separated contributions of the $CC$, $T$, and $T'$
responses are shown (the $CL$ and $LL$ components are not displayed
because they are smaller). For forward angles the $CC$ response is
larger than $T'$, but the $T$ response dominates. This arises because
in neutrino scattering the axial current contributes significantly to
$T$, while its role in $CC$ is minor. With increasing angle the $CC$
contribution decreases relative to $T$ and $T'$, and the $T'$ component
becomes more important. In summary, except for very small angles, the
neutrino cross section is dominated by the transverse response, which
explains why the relative impact of MEC does not vary strongly with
kinematics. A more detailed study would be required to quantify in
depth the kinematic dependence of MEC effects.

%--------------------------
\section{Discussion and Concluding Remarks}
%--------------------------

In this work we have presented a detailed study of meson-exchange
currents in the 1p1h channel of CC neutrino and antineutrino
scattering on nuclei. Our analysis focused on the interference between
one-body and two-body currents, an effect often neglected in neutrino
event generators but that, as we have shown, can be sizable and may
contribute to the systematic uncertainties in oscillation experiments.

The formalism employed was developed starting from the relativistic
Fermi gas model, including 1p1h excitations produced by one-body and
two-body current operators.  In this context, we computed the five
nuclear response functions ($CC$, $CL$, $LL$, $T$, and $T'$) including
the interference between the 1b current and the seagull,
pion-in-flight, pion-pole and $\Delta$ currents.  The non-relativistic
limit of the response functions has been carefully examined as a
preliminary step to validate the relativistic model; in this limit,
the spin sums can be calculated analytically, allowing us to
understand the signs and relative importance of the dominant terms.
In particular, we
found that the $\Delta$ current gives rise to a strong negative
interference in the transverse responses $T$ nad $T'$. This is a new
and important result, which is absent in existing neutrino models and
calls for a careful re-evaluation of the role of 1b–2b interference
terms in neutrino-nucleus scattering.

To assess the model dependence of the interference effects, we
extended the analysis beyond the RFG by employing two additional
nuclear models: the relativistic mean field  in nuclear matter,
and the superscaling approach with effective mass, SuSAM*. In the RMF,
the interaction of the nucleons with scalar and vector mean fields
modifies their dispersion relation through an effective mass $m_N^*$ and
a vector energy shift. These modifications affect the response
functions by shifting their peak positions and redistributing
strength, thereby mimicking to some extent the effects of binding and
final-state interactions in a more realistic manner.

The SuSAM* model, on the other hand, incorporates the relativistic
effective mass from RMF into the superscaling framework. It assumes a
factorization of the nuclear response into a single-nucleon response
multiplied by a phenomenological scaling function, extracted from
inclusive $(e,e')$ data.  One of the advantages of the SuSAM* approach
is that it effectively incorporates nuclear effects beyond the Fermi
gas model, such as final-state interactions (FSI) and other complex
mechanisms, which are phenomenologically embedded on average in the
experimental scaling function.  The SuSAM* model also enables a
consistent inclusion of MEC effects in both the scaling function and
the averaged single-nucleon responses.

Our results show that the 1b–2b interference contributions, though
relatively small compared to the full cross section, are not
negligible and may be comparable in size to other nuclear effects
usually included in more sophisticated models. For instance, effects
such as long-range correlations modeled via the Random Phase
Approximation ~\cite{Nieves:2004wx,Martini2009},
final-state interactions (FSI) in energy-dependent RMF
approaches~\cite{GonzalezJimenez:2017gcy}, finite-size shell effects
in Hartree–Fock based models ~\cite{Jachowicz:2002rr},
or GiBUU transport
theory~\cite{Mos16}, all modify the shape and size of the
nuclear response. The presence of a non-negligible 1b–2b interference
implies that those models may need to be revisited or extended to
estimate at least the order of magnitude of this contribution and its
possible interplay with the aforementioned mechanisms.

The transverse enhancement observed in calculations using correlated
wave functions~\cite{Fab97} or Green’s Function Monte
Carlo~\cite{Lovato2016} may point to nuclear dynamics
beyond the mean-field approximation, possibly involving spin-entangled
two-body structures that interact with the MEC operator but are
averaged out in mean-field models. This may help explain the
enhancement seen in electron scattering data~\cite{Bodek:2011ps},
although in GFMC calculations it is not possible to separate 1p1h and
2p2h final states. This problem requires a shift in perspective that
is currently under investigation.

In view of the present study, it is pertinent to comment on recent
results reported in Refs.~\cite{Fra23,Fra25,Lov23}, where a positive
$T$ interference between one- and two-body currents in electron
scattering has been obtained in the RMF and spectral function
models. This outcome contrasts with our comparative analyses across
several models of ref. \cite{Cas25}, where the interference has
consistently been found to be negative. In particular, the positive
result of Refs.~\cite{Fra23,Fra25} is due to a $\Delta$ current with a
sign opposite to the standard operator, as was explicitly demonstrated
in the Appendix G of Ref.~\cite{Cas25}. A similar feature can be
identified in Ref.~\cite{Lov23}, where the propagator of the $\Delta$
resonance is taken with opposite sign. On the other hand,
Ref.~\cite{Lovato2016} employs a fully correlated model with the
standard operator, in which case the enhancement plausibly originates
from the role of correlations, consistent with the findings of
Ref.~\cite{Fab97}.

Although in this paper we have worked with some of the most elementary
nuclear models, this choice was necessary in order to formulate the
problem rigorously and within a fully reproducible framework. We
believe that this strategy provides a solid baseline that will allow
more realistic models to incorporate and test the effects we have
identified. Our hope is that this study serves as a starting point for
a deeper investigation and helps indicate the direction in which
further theoretical efforts should proceed.

%--------------------------
\section{Acknowledgments}
%----------------------------

The work was supported by Grant No. 
   PID2023-147072NB-I00
funded by MICIU/AEI /10.13039/501100011033 and by ERDF/EU;
by Grant No. 
   FQM-225 
funded by 
   Junta de Andalucia;
by Grant    NUCSYS
funded by 
    INFN;
by Grant No. 
    BARM-RILO-24-01
funded by 
    University of Turin;
    by  the ``Planes Complementarios de I+D+i" program (Grant ASFAE/2022/022) 
by 
MICIU with funding from the European Union NextGenerationEU and 
Generalitat Valenciana.

\clearpage

%----------
\appendix
%----------

%-----------------------------------
\section{Isospin Summations in the 1p1h MEC Matrix Element}
\label{appendix:isospin}
%-----------------------------------------------------------

Here we provide the sums over the isospin index \(t_k\) of the
appearing in the 1p1h MEC matrix element. 
We first note that the MEC can be written as
a linear combination of  $\tau^{(1)}_\pm$ and $\tau^{(2)}_\pm$
and $i[ \ntau^{(1)} \times \ntau^{(2)}]_\pm$
\begin{equation}
j_{2b}^\mu= \tau^{(1)}_\pm j_1^\mu+ \tau^{(2)}_\pm j_2^\mu + 
i[ \ntau^{(1)} \times \ntau^{(2)}]_\pm j_3^\mu,
\end{equation}
where the 2b currents $j_1^\mu$,  $j_2^\mu$, and $j_3^\mu$
are isospin-independent. 
To be specific, we will consider the case of the \( (+) \) component
of the isospin current, corresponding to $N \rightarrow P$ transitions.
\begin{equation}
\tau_+= \tau_1 + i\tau_2 =
\begin{pmatrix}
0 & 2 \\
0 & 0
\end{pmatrix},
\end{equation}
\begin{equation}
i[\ntau^{(1)} \times \ntau^{(2)}]_+=
\tau_+^{(1)}\tau_3^{(2)}
-\tau_3^{(1)}\tau_+^{(2)},
\end{equation}
from where we obtain
\begin{eqnarray}
\tau_+|P\rangle = 0,
&&
\tau_+|N\rangle = 2|P\rangle 
\label{tau}
\end{eqnarray}
\begin{eqnarray}
  i[\ntau^{(1)} \times \ntau^{(2)}]_+  |NP\rangle &=& 2|PP\rangle,
  \nonumber\\
i[\ntau^{(1)} \times \ntau^{(2)}]_+  |NN\rangle &=& -2|PN\rangle+2|NP\rangle,
,\nonumber\\
i[\ntau^{(1)} \times \ntau^{(2)}]_+  |PP\rangle &=&0 
\nonumber\\
i[\ntau^{(1)} \times \ntau^{(2)}]_+  |PN\rangle &=&-2|PP\rangle
\label{tautau}
\end{eqnarray}
From these elementary results, it is straightforward
to compute the isopin sums. The results are the following.

\paragraph{Direct terms.}
\begin{equation} \label{iso1}
\sum_{t_k=\pm 1/2} 
\langle P t_k|  \tau^{(1)}_+  | N t_k\rangle
=4,
\end{equation}
\begin{equation} \label{iso2}
\sum_{t_k} 
\langle Pt_k| \tau^{(2)}_z  | Pt_k\rangle
=0,
\end{equation}
\begin{eqnarray}
\sum_{t_k} 
\langle Pt_k| i[\ntau^{(1)} \times \ntau^{(2)}]_+  | Nt_k\rangle 
&=&0. 
\label{iso3}
\end{eqnarray}

\paragraph{Exchange terms.}
\begin{equation} \label{iso4}
\sum_{t_k} 
\langle Pt_k|  \tau^{(1)}_+  | t_kN\rangle
=2,
\end{equation}
\begin{equation} \label{iso5}
\sum_{t_k} 
\langle Pt_k| \tau^{(2)}_+  | t_kN\rangle
=2,
\end{equation}
\begin{eqnarray}
\sum_{t_k} 
\langle Pt_k| i[\ntau^{(1)} \times \ntau^{(2)}]_+  | t_kN\rangle 
=-4. 
 \label{iso6}
\end{eqnarray}
Similar results are obtained for the $P\rightarrow N$ transition with the
$(-)$ isospin components.

After performing these sums the effective one-body matrix element
$j_{2b}(\mathbf{p},\mathbf{h})$ induced by the two-body current, Eq.
(\ref{effectiveOB}), can be written as sum of direct minus exchange currents
\begin{equation}
j_{2b}^{\mu}(\np,\nh) =
j_{2b}^{\mu}(\np,\nh)_{\rm dir}- 
j_{2b}^{\mu}(\np,\nh)_{exch}, 
\end{equation}
  where
\begin{eqnarray}
j_{2b}^{\mu}(\np,\nh)_{\rm dir}
&=&
\frac{1}{V}\sum_\nk\sum_{s_k}
 4j_1^{\mu}(\np,\nk,\nh,\nk)
\\
j_{2b}^{\mu}(\np,\nh)_{\rm exch}
&=&
\frac{1}{V}\sum_\nk\sum_{s_k}
\left[
2j_1^{\mu}(\np,\nk,\nk,\nh)
\right.
\nonumber\\
&& \kern -1cm
\left.
+2j_2^{\mu}(\np,\nk,\nk,\nh)
   -4j_3^{\mu}(\np,\nk,\nk,\nh)
  \right].
\label{exchange}
\end{eqnarray}

%----------------------------------------------------------------------
\section{Relation btween CC and EM interference responses
  in the vector sector}
%------------------------------------------------------------------------
\label{appendix:CC-EM}

In this appendix, we demonstrate that the interference
responses between the one-body and two-body currents in CC neutrino
scattering in the vector sector are exactly twice the corresponding
electromagnetic responses when both proton and neutron  emission are
summed. This relation holds in symmetric nuclear matter and follows
from the isovector nature of both the CC and the electromagnetic
two-body currents.

 Indeed, the one-body
electromagnetic and CC vector currents written as isospin operators are
\begin{eqnarray}
  j_{1b,e}^\mu &=& s^\mu + f^\mu \tau_z \\
  j_{1b,V}^\mu &=& f^\mu \tau_+
\end{eqnarray}
where $s^\mu$ and $f^\mu$ are the isoscalar and isovector 1b currents,
respectively.

Therefore the electromagnetic current in a 1p1h excitation
between isospin states $|t_h\rangle$ and $|t_p\rangle$ is
\begin{equation}
  \langle t_p| j_{1b,e}^\mu(\np,\nh)|t_h\rangle
  = \delta_{t_pt_h}[s^\mu(\np,\nh) + 2t_hf^\mu(\np,\nh)],
\end{equation}
and the CC vector current in a neutron-to-proton transition is
\begin{equation}
  \langle P|  j_{1b,V}^\mu(\np,\nh) |N\rangle= 2f^\mu(\np,\nh).
\end{equation}

The two-body MEC currents are written in a similar way
\begin{eqnarray}
  j_{2b,V}^\mu
  &=& \tau^{(1)}_\pm j_1^\mu+ \tau^{(2)}_\pm j_2^\mu + 
i[ \ntau^{(1)} \times \ntau^{(2)}]_\pm j_3^\mu,
\\
j_{2b,e}^\mu
  &=& \tau^{(1)}_z j_1^\mu+ \tau^{(2)}_z j_2^\mu + 
i[ \ntau^{(1)} \times \ntau^{(2)}]_z j_3^\mu,
\end{eqnarray}
where the 2b vector currents $j_1^\mu$, $j_2^\mu$, and $j_3^\mu$ are
isospin-independent.  Since the two-body current \( (j_{2b})_V \) has
no axial component, the corresponding effective one-body current
contains only the exchange term, as given in Eq.~(\ref{exchange}).
Thus in a transition where $h$ is neutron and $p$ is proton, we have
\begin{equation}
\langle P |j_{2b,V}^{\mu}(\np,\nh)|N\rangle = -j_{2b}^{\mu}(\np,\nh)_{ex} 
\end{equation}
where $j_{2b}^\mu(\np,\nh)_{exc}$ is given by Eq. (\ref{exchange})
with only the vector part.  On the other hand the electromagnetic
current for a transition from isospin $t_h$ to $t_p$ was computed in
ref. \cite{Cas25}, and is given by
\begin{equation}
\langle t_p|(j_{2b,e}^{\mu})(\np,\nh)|t_h\rangle = -\delta_{t_pt_h}t_h j_{2b}^{\mu}(\np,\nh)_{ex} 
\end{equation}
Let us compute a generic contribution to the interference tensor \(
w_{1b2b}^{\mu\nu} \) of Eq. (\ref{w1b2b}),
such as \( j_{1b}^{\mu *} j_{2b}^{\nu} \), with the
one-body and two-body currents.
In the case of neutrino scattering the vector part is
\begin{equation}
  \langle P|j_{1b,V}^{\mu}|N\rangle ^*
  \langle P|j_{2b,V}^{\nu}|N\rangle = -2f^{\mu *}(j_{2b}^\nu)_{ex}.
\label{V}
\end{equation}
In the em case it is 
\begin{equation}
  \langle t_p|  j_{1b,e}^{\mu} | t_h\rangle ^*
  \langle t_h| j_{2b,e}^{\nu} | t_h\rangle =
  -\delta_{t_pt_h}(s^\mu+2t_hf^{\mu})^*t_h(j_{2b}^\nu)_{ex}.
\end{equation}
Performing the sum over isospin in the em case,
\begin{equation}
\sum_{t_pt_h=\pm1}  \langle t_p|  j_{1b,e}^{\mu} | t_h\rangle ^*
  \langle t_h| j_{2b,e}^{\nu} | t_h\rangle =
  -f^{\mu *}(j_{2b}^\nu)_{ex}.
\label{em}
\end{equation}
This factor of two between the CC (\ref{V}) and EM (\ref{em})
expressions thus applies to all interference response functions
considered in this work.

%################################################################
\section{Non relativistic reduction of the axial $\Delta$ current}
\label{appCbis}
%################################################################

Here we compute the non relativistic axial $\Delta$ current to leading
order in the $1/m_N$ expansion.  The non-relativistic reduction
follows the same procedure as that used for the vector current in
Ref. \cite{Cas25}.  We start by writing the forward and backward
relativistic currents (\ref{deltaF},\ref{deltaB}) in the form
\begin{eqnarray}
  j^\mu_{\Delta F}= U_F(1,2) K_F^\mu + (1\leftrightarrow 2), \\ 
  j^\mu_{\Delta B}= U_B(1,2) K_B^\mu + (1\leftrightarrow 2)
\end{eqnarray}
where
\begin{eqnarray}
  K_F^\mu &=& \frac{ff^*}{m_\pi^2}C_5^A V(2'2) A^\mu, \label{KF}\\
  K_B^\mu &=& \frac{ff^*}{m_\pi^2}C_5^A V(2',2) B^\mu, \label{KB}
\end{eqnarray}
  and we have defined
\begin{eqnarray}
  A^\mu  &=& \bar{u}(1')k_{2}^\alpha G_{\alpha \beta}(p_{1}+Q)g^{\beta\mu}u(1),
\\
  B^\mu  &=&
  \bar{u}(1')k_{2}^\beta g^{\mu\alpha}G_{\alpha\beta}(p'_{1}-Q)u(1).
\end{eqnarray}
In the following, we omit the explicit spinors, $u(1)$,  $u(1')$,
and reduce the
expressions to leading order.
Using $k_2^\alpha \rightarrow (0,\nk_2)$, we have
\begin{eqnarray}
  A^\mu &=&      k_{2}^\alpha G_{\alpha \beta}g^{\beta\mu}
  \rightarrow k_{2}^k     G_{k \beta}    g^{\beta\mu}.
\label{amu} \\
  B^\mu &=&      k_{2}^\beta g^{\mu\alpha} G_{\alpha \beta}
  \rightarrow k_{2}^k  g^{\mu\alpha}    G_{\alpha k}   
\label{bmu}
\end{eqnarray}

\subsubsection{Time components}
%------------------------------

We first show that the time component of the axial $\Delta$ current,
vanishes in the static limit. In fact the forward and backward parts are,
respectively, proportional to
\begin{eqnarray}
  A^0  &\rightarrow& k_{2}^{k}G_{k\beta}g^{\beta 0}  =k_{2}^{k}G_{k0}g^{0 0}, \\
  B^0  &\rightarrow& k_{2}^{k}g^{\alpha 0}G_{\alpha k} = k_{2}^{k}g^{0 0}G_{0 k}.
\end{eqnarray}
Writing the  $\Delta$ propagator in the static limit in the form \cite{Cas25}
\begin{equation}
  G_{\alpha\beta} \sim -
  \frac{
   g_{\alpha\beta}-\frac{\gamma_\alpha\gamma_\beta}{3}
  -\frac{2P_{\alpha}P_{\beta}}{3m_\Delta^2}
  +\frac{P_{\alpha}\gamma_{\beta}-P_{\beta}\gamma_{\alpha}}{3m_\Delta}
   }{m_N-m_{\Delta}},
\end{equation}
we have
\begin{equation}
  G_{k0} \sim -
  \frac{
   g_{k0}-\frac{\gamma_k\gamma_0}{3}
  -\frac{2P_kP_0}{3m_\Delta^2}
  +\frac{P_k\gamma_0-P_0\gamma_k}{3m_\Delta}
   }{m_N-m_{\Delta}}.
\end{equation}
In the static limit $P_k \rightarrow 0$, and $\gamma_k\rightarrow 0$. Then
$G_{k0}\rightarrow 0$
and consequently $A^0\rightarrow 0$. Similarly we find $G_{0k}\rightarrow 0$
and $B^0\rightarrow 0$.

\subsubsection{Space components}
%------------------------------------------

Using (\ref{gammas2}), the spatial components $G_{kj}$ of the $\Delta$
propagator to leading order are written as \cite{Cas25},
\begin{eqnarray}
  G_{kj} &\sim& 
  \frac{  -1 }{m_N-m_{\Delta}}
  \left(  g_{kj}-\frac{\gamma_k\gamma_j}{3} \right)
\nonumber\\
&\rightarrow&
  \frac{  1 }{m_N-m_{\Delta}}
  \left( \frac23  \delta_{kj}-\frac{i}3 \epsilon_{kjl}\sigma_l \right).
\end{eqnarray}
Substituting into Eq. (\ref{amu}) we obtain
\begin{eqnarray}
  A^i &\sim & k_{2}^k     G_{k \beta}    g^{\beta i}
            = k_{2}^k     G_{k j}    g^{j i} \nonumber\\
&\rightarrow&
 k_2^k \frac{  1 }{m_N-m_{\Delta}}
  \left(\frac23  \delta_{kj}-\frac{i}3 \epsilon_{kjl}\sigma_l \right) g^{ji}.
  \nonumber\\
  &=&
  \frac{  1 }{m_\Delta-m_N}
  \left( \frac23 k_2^i-\frac{i}3 \epsilon_{kil}k_2^k\sigma_l \right).
\end{eqnarray}
Or, in vector form
\begin{equation}
\nA  \rightarrow
  \frac{  1 }{m_\Delta-m_N}
  \left( \frac23 \nk_2 +\frac{i}3\nk_2\times \nsigma_l^{(1)} \right).
\end{equation}
(recall that the operator $A^\mu$ was acting on the spinor $u(1)$).
Similarly, from Eq. (\ref{bmu}),
\begin{eqnarray}
  B^i &\sim & k_{2}^k     g^{\alpha i}    G_{\alpha k}
            = k_{2}^k     g^{ji}    G_{j k} \nonumber\\
&\rightarrow&
 k_2^k  g^{ji} \frac{  1 }{m_N-m_{\Delta}}
  \left(\frac23  \delta_{jk}-\frac{i}3 \epsilon_{jkl}\sigma_l \right)
  \nonumber\\
  &=&
  \frac{  1 }{m_\Delta-m_N}
  \left( \frac23 k_2^i-\frac{i}3 \epsilon_{ikl}k_2^k\sigma_l \right).
\end{eqnarray}
Moreover, in vector notation, $\nB$ can be expressed as:
\begin{equation}
\nB  \rightarrow
  \frac{  1 }{m_\Delta-m_N}
  \left( \frac23 \nk_2 -\frac{i}3\nk_2\times \nsigma_l^{(1)} \right).
\end{equation}
Once we have the non-relativistic expressions of $\nA$ and $\nB$, we
can insert them into the definitions of $\nK_F$ and $\nK_B$, given in
equations (\ref{KF}) and (\ref{KB}). Applying also the
non-relativistic limit of the function $V$ from equation (\ref{Vnorel}), we have
\begin{eqnarray}
  \nK_F &\rightarrow& -\frac{ff^*}{m_\pi^2}C_5^A
  \frac{\nk_2\cdot\nsigma^{(2)}}{\nk_2^2+m_{\pi}^2}
  \nA,
  \\
  \nK_B &\rightarrow& -\frac{ff^*}{m_\pi^2}C_5^A
  \frac{\nk_2\cdot\nsigma^{(2)}}{\nk_2^2+m_{\pi}^2}
  \nB.
\end{eqnarray}

\subsubsection{Total axial $\Delta$ Current}

Using Eqs. (\ref{UF}) and (\ref{UB}) for the isospin operators $U_F$
and $U_B$ we can write the axial $\Delta$ current in the form
\begin{equation}
  \nj_{\Delta A}= \frac{2}{\sqrt{6}}\tau_+^{(2)}(\nk_F+\nK_B)
    +\frac{1}{\sqrt{6}}i[\ntau^{(1)}\times\ntau^{(2)}]_+(\nk_B-\nK_F)
\end{equation}
To leading order we have
\begin{eqnarray}
  \nA+\nB &=& \frac{1}{m_\Delta-m_N}\frac43 \nk_2, \\
  \nB-\nA &=& -\frac{1}{m_\Delta-m_N}\frac{2i}{3} \nk_2\times\nsigma^{(1)}.
\end{eqnarray}
Thus, we obtain
\begin{eqnarray}
  \nK_F+\nK_B &=& -\frac{ff^*}{m_\pi^2}\frac{C_5^A}{m_\Delta-m_N}
  \frac{\nk_2\cdot\nsigma^{(2)}}{\nk_2^2+m_{\pi}^2}\frac43\nk_2,
  \\
  \nK_B-\nK_F &=& \frac{ff^*}{m_\pi^2}\frac{C_5^A}{m_\Delta-m_N}
  \frac{\nk_2\cdot\nsigma^{(2)}}{\nk_2^2+m_{\pi}^2}
  \frac{2i}{3} \nk_2\times\nsigma^{(1)}.
  \nonumber\\
\end{eqnarray}
Finally, the non-relativistic axial $\Delta$ current takes the form:
\begin{eqnarray}
  \nj_{\Delta A}&=&
  - \sqrt{\frac{3}{2}}\frac29
    \frac{ff^*}{m_\pi^2}\frac{C_5^A}{m_{\Delta}-m_N}
   \left[
     4\tau_+^{(2)} \frac{\nk_2\cdot\nsigma^{(2)}}{\nk_2^2+m_{\pi}^2}\nk_2
     \right.
     \nonumber\\
&& \left.    + [\ntau^{(1)}\times\ntau^{(2)}]_{+}
  \frac{ \nk_2\cdot\nsigma^{(2)} }{ \nk_2^2+m_{\pi}^2 } \nk_2\times\nsigma^{(1)}
  \right]
\nonumber\\
&&   + (1\leftrightarrow2),
\end{eqnarray}
By writing explicitly the $(1 \leftrightarrow 2)$ term, we directly arrive at
Eq. (\ref{deltaAxial}), which matches the form given in Ref.~\cite{Ris89}.

%%%%%%%%%%%%%%%%%%%%%%%%%%%%%%%%%%%%%%%%%%%%%%%%%%%%%%%%%%%%%%%%%%%

\end{document}